\documentclass[12pt]{article}
\usepackage{amsmath}
\usepackage{bm}
\usepackage{amssymb}
\usepackage{amsthm}
\usepackage{graphicx,psfrag,epsf}
\usepackage{subcaption}
\usepackage{enumerate}
\usepackage[authoryear]{natbib}
\usepackage{url} 
\usepackage{authblk}
\usepackage{float}
\usepackage{algorithmic}
\usepackage[ruled,vlined]{algorithm2e}
\usepackage{hyperref}
\usepackage{cleveref}
\usepackage{booktabs} 
\usepackage{tabularx}
\usepackage{graphicx}
\usepackage{xcolor}
\usepackage{comment}
\usepackage{booktabs}

\graphicspath{{fig/}}

\newtheorem{remark}{Remark}

\usepackage
[acronym,nowarn,section,nogroupskip,nonumberlist]{glossaries}
\setacronymstyle{long-sc-short}
\glsdisablehyper{}
\setglossarystyle{long3col}
\makeglossaries
\newacronym{method}{dbml}{Distributional Balancing using Machine Learning}

\DeclareMathOperator*{\argmin}{argmin}

\begin{document}

\def\spacingset#1{\renewcommand{\baselinestretch}%
{#1}\small\normalsize} \spacingset{1}


  \title{Distributional Balancing with Machine Learning for Clinical Trial Augmentation Using Real-World Data}
  \renewcommand\Authands{, }
  \author[1]{Zern Ke}
  \author[1]{Mingshi Cui}
  \author[1]{Gemma Moran}
  \author[1]{Javier Cabrera}
  \affil[1]{Department of Statistics, Rutgers University, New Brunswick}
  \maketitle

\bigskip
\begin{abstract}
In clinical trials, randomization of treatment and control groups is typically used to ensure the groups have similar covariate distributions on average, resulting in unbiased causal effect estimation. Such balanced covariate distributions are hard to achieve in practice, however, due to recruitment costs, patient dropouts, and more. One possible solution to this problem is to include control patients from external, real world databases. 
 In this paper, we propose \gls{method}, a method that selects control units from a real world database by matching the distribution between the treatment group and the potential control group, instead of matching units between two groups. \gls{method} has three steps. First, we detect anomalous database units (with respect to the treatment distribution) using a variational autoencoder. Second, we re-weight the remaining database units to match the distribution of the treatments. Finally, we use these weights to sample units for our control group.
The proposed method is compared to alternative matching-based algorithms and weighted algorithms,  achieving superior performance in covariate balance. 
\end{abstract}

\noindent%
\vfill

\newpage
\spacingset{1.8} 

\section{Introduction}
\label{sec:intro}

\emph{Randomized controlled trials (RCTs)} are the gold standard to estimate causal effects in clinical and epidemiological studies. Randomization ensures that treatment assignment is independent of observed and unobserved confounders, thereby balancing patient characteristics between treatment groups in expectation and allowing causal effects to be estimated \citep{fisher1935, rubin1974, rosenbaum2002}. However, randomization is not always feasible.  For example, in the case of {rare diseases}, the small number of available patients makes it difficult to recruit enough participants for a fully randomized study. In other cases, {ethical considerations} prevent randomization of patients to a placebo or control group when experimental therapy shows early evidence of substantial benefit. In such situations, researchers would like to take advantage of large external databases such as electronic health records, insurance claims data and disease registries. These data can be used as external controls to supplement or replace randomized control arms. External controls can also ameliorate practical issues with clinical trials, such as limited time, or regulatory constraints. As a bonus, the use of the external control arm could also dramatically reduce the cost and time of the trial \citep{hernanrobins2020, fda2022, ema2021}.  

A key challenge in selecting external control data is to ensure balance between the characteristics of treated and control patients. This challenge is particularly difficult in clinical and epidemiological research where many patient characteristics are categorical (e.g. presence/absence of comorbidities, risk factors) and ordinal (e.g. disease severity, risk stratification), as well as continuous (e.g. labratory values). 

A classic approach to obtaining controls is matching; matching has been leveraged by \citet{yu2020matching} to obtain external control units. Common implementations of matching include exact matching, nearest-neighbor matching, and propensity score matching \citep{rosenbaum1983, stuart2010, ventz2019}.  

There are several challenges with applying conventional matching methods in this setting. Matching requires a well-defined notion of similarity between treated and control units, but this becomes difficult when the covariates contain a mixture of binary, categorical, ordinal, and continuous variables. Although distance-based matching can be effective for continuous covariates, it may not adequately capture the joint dependence structure among mixed-type variables. Conversely, exact or coarsened matching on categorical variables can become too restrictive when the number of covariates is large, because even small differences across several discrete variables may leave many treated units without close control counterparts \citep{stuart2010, austin2014, wang2017flame}. Therefore, in this setting, the main limitation is not only marginal covariate imbalance, but also the difficulty of defining and achieving similarity at the level of the joint covariate distribution.

Second, matching methods can struggle in high-dimensions. For nearest-neighbor matching, in high-dimensions, the notion of ``nearest neighbor'' loses interpretability: most units appear equally distant from each other, making it difficult to identify genuinely similar matches \citep{beyer1999, abadie2006}. Propensity-score matching reduces high-dimensional covariates to a one-dimensional sufficient summary, but can underperform when the propensity score model is misspecified \citep{Drake1993PSMisspec}.  In addition, for high-dimensional data, propensity estimation can be noisy; units with similar propensity scores may still differ systematically in the original covariate space \citep{SmithTodd2001ReconcilingPSM, DiamondSekhon2013GenMatch, Hainmueller2012EntropyBalancing}. 

Third, matching methods require specifying threshold parameters, such as a caliper, that defines how close two units must be to be considered neighbors. A poor parameter choice will result in either no match, causing treated units to be dropped from the analysis, or poor-quality matches that do not adequately balance confounders between groups \citep{rosenbaum2002}. 

\begin{figure}[H]
    \centering
    \includegraphics[width=\linewidth]{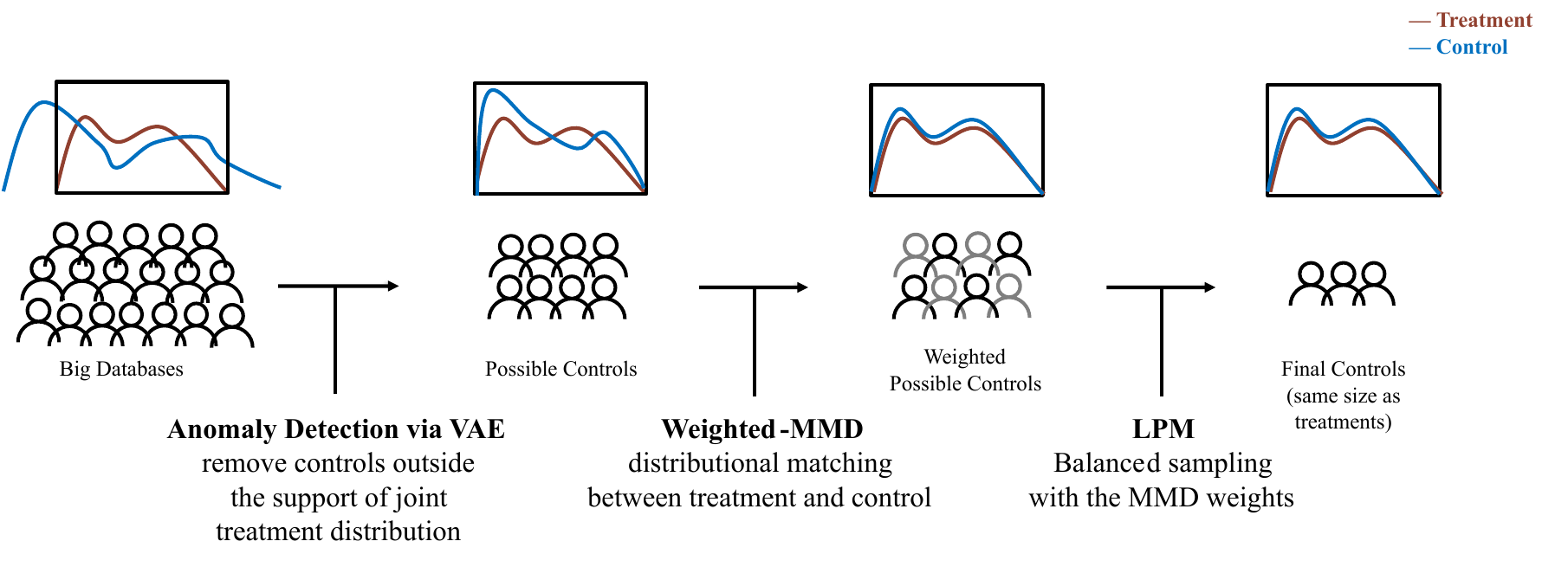}
    \caption{\gls{method} Pipeline}
    \label{fig:method overview}
\end{figure}

Instead of matching treated and external control units one-to-one or one-to-many, we consider a {distributional balancing} approach. That is, the selected controls are chosen such that their joint covariate distribution is close to the given treatment group.  To accomplish this, we propose \gls{method}. \gls{method} has three steps (Figure \ref{fig:method overview}).
\begin{itemize}
\item First, we detect and omit out-of-distribution control samples using a variant of the two-stage variational autoencoder (VAE), which is well-suited for covariates of different data-types \citep{ma2020vaem}. The remaining control samples are then designated as ``possible control samples''.  
\item Second, we assign each possible control a learnable weight; we then optimize the weights to minimize the Maximum Mean Discrepancy \citep[MMD,][]{mmd-gretton12a} between the weighted possible controls and treatment group. The MMD is a measure of distance between distributions; this ensures the weighted possible controls and treatment group are close in distribution. 
\item Third, we construct our final control group by sampling from the possible control group based on the learned weights. In particular, the sampling strategy we use is the local pivotal method \citep[LPM,][]{lpm-Friedman,lpm-Grafstrom} with the MMD weights as the initial inclusion probability.  The LPM is a sampling algorithm that sequentially adjusts inclusion probabilities between nearby units to achieve a balanced sample across covariate-space.
\end{itemize}

Our \gls{method} can also be readily extended to settings where the clinical dataset contains both treated patients and existing controls. In this case, additional controls can be selected from an external candidate pool to construct a combined control group, whose covariate distribution is well balanced with the treated group. For simplicity, throughout the main paper, we focus on the setting where the target group contains only treated patients. Extensions to settings with existing controls are discussed in Section~\ref{sec:diss}.

We evaluate \gls{method} using synthetic, semi-synthetic, and real-world datasets. The results in Section~\ref{sec:exp} demonstrate that \gls{method} achieves generally superior covariate balance between treated and control units relative to alternative methods. Our implementation is available at \url{https://github.com/Zernjk/DBML}.

\subsection{Related work}

Methods that leverage historical individual-level data or external summary data to improve inference on treatment effects are widely used in areas such as adaptive designs and case–control studies. Broadly, methods fall in two categories: parameter-focused methods, which leverage past data to inform parameter estimation; and control group augmentation, which add past data to the current control group. Our proposed method  augments the control group; we also review parameter-focused methods for completeness. 

\textbf{Parameter-focused methods.} 
 Both Bayesian and frequentist approaches have been developed to leverage past data to inform model parameter estimation. For example, under Bayesian frameworks, historical data are often incorporated through priors on the parameters of the outcome model, denoted by $\theta$, such as the power prior. The power prior incorporates historical data $D_0$ as: $ \pi(\theta \mid D_0, a_0) \propto \mathcal{L}(D_0 \mid \theta)^{a_0}\,\pi_0(\theta)$, where  $a_0\in[0,1]$ is a weight which adjusts how much the historical data $D_0$ contributes to posterior inference \citep{psioda2018historical, IbrahimChen2000}.

Meanwhile, frequentist approaches such as empirical-likelihood and constrained-likelihood encode external summary information as estimating-equation constraints  and combine these with the likelihood of the internal study, solving the resulting constrained likelihood via Lagrange multipliers or profiling to gain efficiency over internal-only MLE \citep{qin2015covariateprevalence,chatterjee2016cml}.

\textbf{Control group augmentation.} Another category of approaches  select subsets of an external big database to augment (or comprise) the control group. A classic approach to obtaining controls is matching; matching has been leveraged by \citet{yu2020matching} to obtain external control units.

Most similar to our proposed method are methods  which determine control groups based on a overall distribution balance criterion (as opposed to one-to-one or one-to-many matches). These papers assign to each control sample $x_i$ a learnable weight $w_i$; the objective is to learn weights such that re-weighted control group has the same covariate distribution as the treatment group. 

\citet{Hainmueller2012EntropyBalancing} propose to minimize the KL-divergence between the weighted control samples and a chosen ``base measure'', with the constraint that the weighted control samples have the same statistics as the treatment group. That is \citet{Hainmueller2012EntropyBalancing} optimize for $W=(w_i)_{i=1}^n$ in the objective:
\begin{align}
    \min_{W}\;\; \sum_{i\in \mathrm{database}} w_i \log\!\Big(\frac{w_i}{q_i}\Big)
    \quad\text{s.t.}\quad
    C W = M,\ \ \sum_{i\in \mathrm{database}} w_i = 1,\ \ w_i>0
    \label{eq:hain_entropy}
\end{align}
where $q_i$ are the chosen base weights, and
 $C$ is the constraint matrix where the components $c(X_i) \;=\; \big(c_1(X_i),\, \ldots,\, c_R(X_i)\big)^{\top}$  are features to balance (e.g. means, variances). Meanwhile, $M \;=\; (m_1,\,\ldots,\,m_R)^{\top}$ are the corresponding treated group sample features. When $q_i$ is uniform ($q_i=1/n$) this objective maximizes the entropy of the $w_i$ (subject to sample statistic constraints). 
 
More recently, \citet{gao2024improving} derived an efficient influence function for treatment effect estimation with external controls. Similarly to \citet{Hainmueller2012EntropyBalancing}, their proposed treatment effect estimator involves a learnable weight $w_i$ for each control unit; the $w_i$ are learned by maximizing the entropy of the weights, with the additional constraint that the integral probability metric between the weighted controls and treatment group is the same:
$\sum_{i\in \mathrm{database}} w_ig(\bm{x}_i) = \sum_{i \in\mathrm{study}} g(\bm{x}_i)$ for any function $g$.

As an alternative to entropy maximization with constraints, \cite{HulingMak+2024} introduced energy balancing weights (EBW). EBW learns weights which minimize the energy distance \citep{szekely2013energy} between a weighted subsample (e.g. a treatment group) and the entire sample. \citet{HulingMak+2024} also introduced the improved EBW, which additionally  minimizes the imbalance between weighted treated and control arms.

\cite{santra2026distributional} proposed a nonparametric causal inference framework based on the characteristic function distance (CFD) to balance covariate distributions. They show that the maximum mean discrepancy and energy distance are special cases of the CFD, 
and extend the approach to instrumental variable settings for estimating local average treatment effects. 

In our method, we also learn weights for each control sample to ensure balanced distributions; in particular, our second step is to learn weights which minimize the Maximum-Mean Discrepancy \citep[MMD,][]{mmd-gretton12a} between the weighted controls and treatment group. Our main contribution is the three-step pipeline: empirically, we find that the MMD with Euclidean kernel, combined with anomaly detection and LPM subsampling, improves distributional balance relative to previous methods. 

\glsresetall

\section{Methods}
\label{sec:meth}

In this section, we introduce \gls{method}, which consists of three steps. Step 1 (Section \ref{sec:vae}) is the detection and removal of control samples that are out-of-distribution with respect to the treatment. Step 2 (Section \ref{sec:mmd}) is the re-weighting of control samples to match the distribution of the treatment. Step 3 (Section \ref{sec:lpm}) is the covariate-aware sampling of controls to achieve a balanced control group. 

\subsection{Anomaly Detection via Variational Autoencoders}
\label{sec:vae}

\subsubsection{Background}
Our approach is inspired by methods for anomaly detection using generative models.  The goal of anomaly detection is to identify observations that deviate from the typical data distribution.  Classical anomaly detection methods can be effective in low-dimensional settings, but they may struggle when the data contain complex nonlinear dependencies among many variables \citep{chandola2009anomaly, chalapathy2019deep, pang2021deep}. 

Deep generative models, especially variational autoencoders (VAEs) \citep{kingma2014auto, rezende2014stochastic}, provide a flexible framework for learning such nonlinear structure in an unsupervised way. VAEs are probabilistic latent-variable models which approximate the observed data density by learning lower-dimensional representations of the data. 

In anomaly detection, a VAE is trained on typical samples to approximate their distribution. Potential anomalous observations are then scored based on this fitted VAE -- for example, using the data reconstruction loss, among other choices \citep{sakurada2014anomaly, an2015variational, zimmerer2019context, xiao2020likelihood, angiulli2023latentout, chalapathy2019deep}.
Observations with poor scores are then categorized as anomalies.

\subsubsection{Two-Stage VAE for Anomaly Detection}

In tabular data, a challenging problem for anomaly detection is that there are mixed-data types: continuous, binary, categorical, ordinal, or count-valued. Therefore, a deep generative model with a Gaussian likelihood is inappropriate. \citet{nazabal2020handling} addressed this issue through the heterogeneous incomplete VAE (HI-VAE), which uses data-type specific likelihoods. 
Although HI-VAE effectively accommodates mixed-data types, differences in scales and loss functions between data types can lead to imbalanced training. To avoid this imbalance, \citet{ma2020vaem} proposed the VAEM, which first fits separate marginal VAEs for each single variable, and then combines the resulting latent variables into a second, common VAE to capture their joint distribution. 

\begin{figure}[H]
    \centering
    \includegraphics[width=0.8\linewidth]{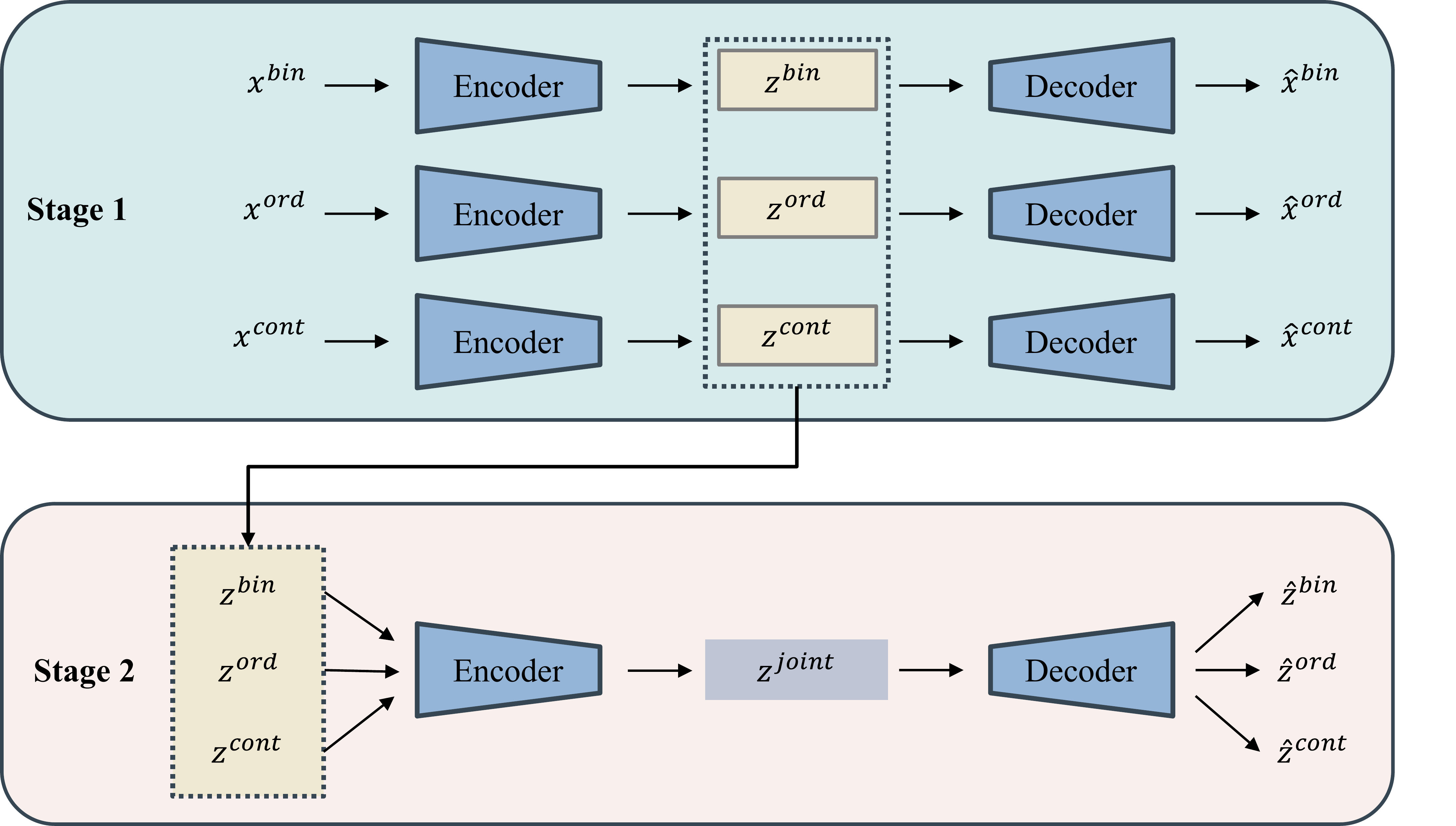}
    \caption{Two-stage Variational Autoencoder. In stage 1, each covariate-type is fit using separate VAEs; in stage 2, the latent representations from stage 1 are concatenated; one VAE is fit to this concatenated representation.}
    \label{fig:2stage vae}
\end{figure}

For our procedure, we adapt the VAEM to handle binary, categorical, ordinal and continuous data (Figure~\ref{fig:2stage vae}). Our procedure, a two-stage VAE, differs from VAEM in that we first fit three separate VAEs for each data-type: binary (including categorical), ordinal, and continuous), instead of a separate VAE for each covariate dimension. The three VAEs have different likelihoods, depending on the covariate type.  We use a Gaussian likelihood for continuous covariates, and Bernoulli and multinomial likelihoods for binary and categorical variables, respectively. For the ordinal covariates, we use an ordered probit model \citep{Wooldridge2010}. For full details, see \Cref{app:vae-details}.

In the second stage, the latent representations learned from the first stage are concatenated and treated as the observed data to which a second VAE is fit. Thus, the second VAE captures the joint distribution of the latent representations across all data-types.

\begin{figure}[H]
    \centering
    \includegraphics[width=0.8\linewidth]{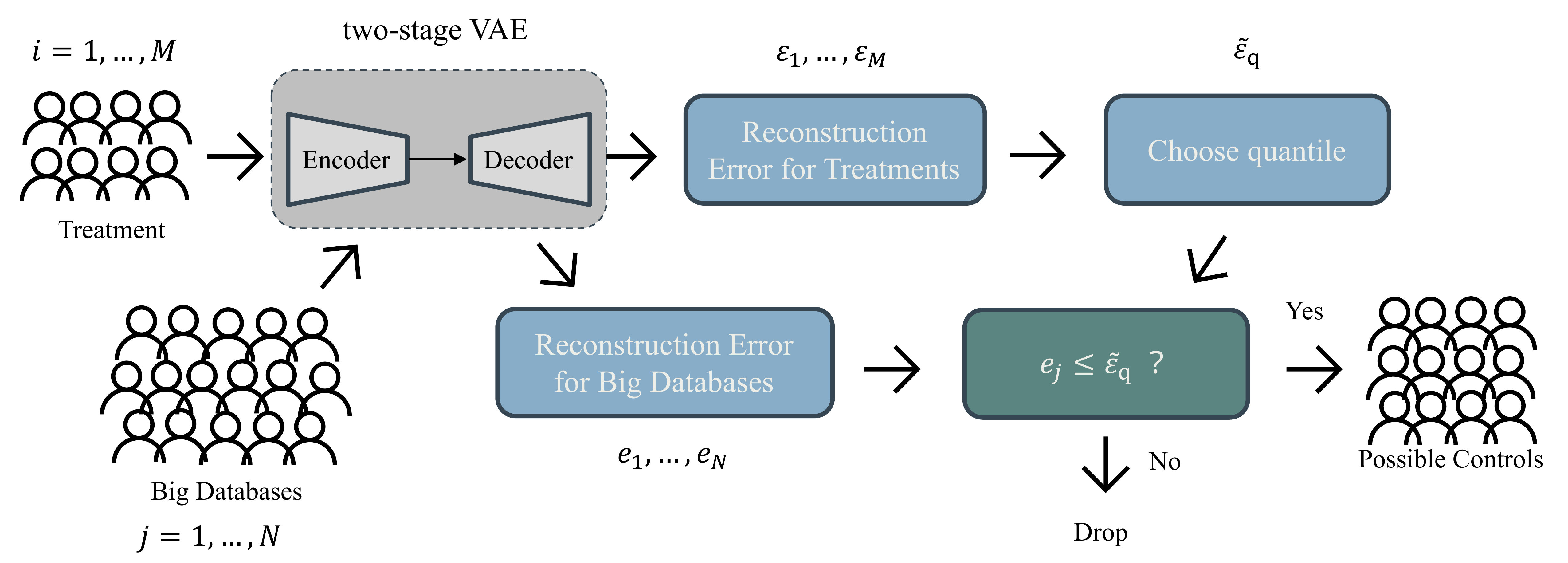}
    \caption{Anomaly Detection using the Two-Stage VAE}
    \label{fig:ad vae}
\end{figure}

After fitting the two-stage VAE to the treated data, the model provides four reconstruction losses to be used in the subsequent anomaly detection process (three data-type specific losses from stage one, and one joint loss from stage two). 
A quantile threshold $q$ is used to identify anomalies. Instances with reconstruction errors above the $q$-th quantile, $\tilde{\varepsilon}_q$, of the treatment-group reconstruction errors are removed, while the rest are retained.
To detect anomalies in both the type-specific and joint distributions, we take the intersection of the retained groups obtained from the four reconstruction losses.

This step excludes controls that are unlikely to be comparable to the treated units before the balancing stage, thereby reducing the burden on the weighting procedure in the next step. In practice, the proposed approach may be especially useful when investigators have access to a large external control database but expect substantial heterogeneity between the database population and the target treated group. 

\subsection{Kernel Maximum Mean Discrepancy Matching}
\label{sec:mmd}
In this section, we describe the second step of the \gls{method} pipeline, where the goal is to learn weights that balance the treated and possible control distributions. Specifically, we learn unit-level weights for the possible controls which minimize the  Maximum Mean Discrepancy \citep[MMD,][]{mmd-gretton12a} between the weighted empirical distribution of the possible controls and the empirical distribution of the treated group.

The MMD measures the discrepancy between two probability distributions. The core idea is to embed each distribution into a reproducing kernel Hilbert space (RKHS) and measure the distance between the resulting mean embeddings. Let $P$ and $Q$ be probability distributions on a common covariate space $\mathcal{X}$, let $\mathcal{H}$ be an RKHS with reproducing kernel $k: \mathcal{X}\times \mathcal{X}\to \mathbb{R}$, and let $\mathcal F$ be the unit ball of $\mathcal{H}$. Given observations $X_1,\dots, X_m\sim P$ and $Y_1,\dots, Y_n\sim Q$,  \cite{mmd-gretton12a} provide the following biased V-statistic estimator of the squared population MMD:
\begin{align}
    \text{MMD}^2_{\mathcal{F},b}(X, Y) = &\frac{1}{m^2} \sum ^m_{i=1} \sum^m_{j =1} k(X_i, X_j) + \frac{1}{n^2} \sum^n_{i=1} \sum^n_{j =1} k(Y_i, Y_j) \notag \\
    &- \frac{2}{mn} \sum^m_{i=1} \sum^n_{j=1}k(X_i, Y_j). \label{eq:mmd-vstat}
\end{align}

We denote $X_1,\dots, X_m\sim P$ as samples from the treated group, and $Y_1,\dots, Y_n$ as samples from the possible control group. We denote the weighted empirical distribution of the possible controls as $\widehat{Q}_\omega = \sum_{j=1}^n \omega_j\delta_{Y_i}/S_\omega$, where $S_\omega = \sum_{j=1}^n \omega_j$ is the total weight. We learn weights $\{\omega\}_{j=1}^n$ which minimize  $\mathrm{MMD}^2_{\mathcal{F},b}$ between the treated group and the weighted empirical distribution of the possible controls:
\begin{align}
    \text{MMD}_{\mathcal{F},b}^2(X, Y, \omega) = &\frac{1}{m^2} \sum ^m_{i=1} \sum^m_{j =1} k(X_i, X_j) + \frac{1}{S_{\omega}^2} \sum^n_{i=1} \sum^n_{j =1} k(Y_i, Y_j)\omega_i\omega_j \notag \\
    &- \frac{2}{mS_{\omega}} \sum^m_{i=1} \sum^n_{j=1}k(X_i, Y_j)\omega_j.
    \label{eq:wmmd-v}
\end{align}
We parameterize $\omega_j =\mathrm{sigmoid}(\beta_j)$ to ensure raw weights are in $(0,1)$.  Minimizing Eq. \ref{eq:wmmd-v} with respect to $\bm{\beta}=(\beta_1,\dots, \beta_n)$, we obtain the MMD weights for each unit in the possible control group. These estimated weights are then used in the next step to guide probability sampling, so that units which better support covariate balance are
more likely to be selected into the final control group.

\begin{remark}
When using MMD as a balancing criterion for weighted empirical distributions, we found the weighted V-statistic form  Eq.~\ref{eq:wmmd-v} to be more effective  than the weighted U-statistic form  Eq.~\ref{eq:mmd-ustat} (the unbiased empirical estimator in \cite{mmd-gretton12a}). This is because the U-statistic induces more sparsity in the weights; for more discussion, see Appendix \ref{sec:mmd-discussion}. 
\end{remark}


\subsection{Local Pivotal Method}
\label{sec:lpm}

In this section, we outline the third step of \gls{method}. Here, we sample from the potential control group using the weights obtained in the previous step in a manner that preserves coverage of the original population space. 

To select a balanced control sample with sample size equal to the treatment group, we use the Local Pivotal Method (LPM) \cite{lpm-Friedman,lpm-Grafstrom}. LPM uses covariate information to guide selection so that the sampled controls are well-spread over covariate-space. The LPM algorithm begins with each unit assigned an inclusion probability; in our case, these are the normalized weights from the MMD step:
\begin{align}
    \pi_i = \frac{\hat{\omega}_i N_{\text{trt}}}{\sum_{i=1}^n \hat{\omega}_i},
\label{eq:inclprob}
\end{align}
where $N_{\mathrm{trt}}$ is the size of the treated group. This normalization ensures the expected sample size is $N_\mathrm{trt}$. Pairs of nearby units, identified dynamically via a distance metric (e.g. Euclidean distance) in the covariate space, are iteratively selected, and their joint inclusion probabilities are updated to either 0 or 1 through a local randomization step that preserves expected inclusion probabilities. Some inclusion probabilities may be greater than one after normalization; we implement a post-processing step to truncate these probabilities (see Appendix \ref{ap:lpm-processing} for more details).

Compared to global balancing algorithms, this local approach requires only limited pairwise calculations in each step, maintaining computational feasibility for large datasets while still ensuring approximate balance with respect to the covariates. The result is a sample that reflects both design weights and the geometric structure of the population covariates, thus improving estimation efficiency.


\glsresetall

\section{Experiments}
\label{sec:exp}

\subsection{Simulation studies}
\label{sec:sim}
In this section, we empirically evaluate \gls{method} on a suite of simulation studies.

\textbf{Metrics.} To evaluate covariate balance, we use the Super-Learner Index (SLI) score \citep{Cabrera2026AC}, defined as the standard deviation of Super-Learner propensity scores estimated using treatment assignment as the response and covariates as predictors. To reduce variability across data splits, we report the average SLI:
\begin{align}
\overline{\mathrm{SLI}} = \frac{1}{M}\sum_{m=1}^M
\mathrm{sd}\left({p_i^{(m)}}_{i=1}^n\right),
\label{eq:sli}
\end{align}
where $m$ indexes the data split. Better covariate balance makes treatment assignment less predictable from the covariates, yielding propensity scores closer to 0.5 and therefore a smaller $\overline{\mathrm{SLI}}$. Thus, lower $\overline{\mathrm{SLI}}$ indicate greater similarity between the treated and control covariate distributions. Additional details are provided in Appendix~\ref{app:sli}.

\textbf{Methods compared.} In the simulation study, we compare to:

\textit{Propensity-score matching} \citep{rosenbaum1983, yu2020matching}: Specifically, we use \texttt{MatchIt} R package \citep{ho2011matchit}, with propensity scores estimated by a generalized linear model $\text{logit}(e(X_i)) = \beta_0 + \beta^T X_i$, and nearest neighbor matching based on the estimated propensity scores $\min_{j \in \text{controls}} \left\vert{} e(X_i) - e(X_j) \right\vert$.

\textit{Entropy balancing} \citep{Hainmueller2012EntropyBalancing}: In our implementation, we balance covariate moments up to the third order $c(X_i) = (X_i, X_i^2, X_i^3)^\top$ in Eq.~\ref{eq:hain_entropy}, thereby requiring agreement in the first three moments of the covariates between the treatment group and the reweighted potential control group. We consider two versions of \textit{Entropy balancing}: 

\begin{itemize}
\item \textit{Entropy}: implements the objective of \cite{Hainmueller2012EntropyBalancing} only, with random sampling of learned weights to obtain the final control group.
\item \textit{DBML-Entropy}: the \gls{method} pipeline, with the MMD step replaced by the entropy balancing objective. 
\end{itemize}

For \gls{method}, we consider two kernels: the RBF kernel $k(X, Y) = \exp (-\frac{||X-Y||^2_2}{2\sigma^2})$ and the Euclidean kernel $k(X, Y)=-||X-Y||_2$. 

\subsection{Synthetic data generation}

We considered four key factors in the simulation design: (i) sample size of treatment group, (ii) covariate correlation, (iii) covariate dimension, and (iv) distributional heterogeneity in the external database. Each factor was varied at two levels, leading to total of $2^4=16$ simulation settings. This design allows us to systematically assess the performance of \gls{method} across low- and high-complexity regimes, including settings with high-dimensional covariates and substantial heterogeneity in the external controls.

In the simulation study, distributional heterogeneity in the external database ($150,000$ units) is induced by generating the external data from a mixture of three distributions, $f_1$, $f_2$, and $f_3$. $f_1$ has the same support and distribution as the target treated  distribution. $f_2$ has the same support as the  target treated distribution but differs in its distributional shape. $f_3$ differs from the treated target distribution both in distributional shape and in support, with only partial support overlap. Therefore, the mixture proportions of $f_1$, $f_2$, and $f_3$ determine the degree of contamination in the external database, where contamination refers to the proportion of external units not generated from the target-aligned distribution $f_1$. The four simulation factors and their corresponding  levels are summarized in Table~\ref{tab:simulation-factors}.

\begin{table}[htbp]
\centering
\normalsize
\caption{Factors varied in the simulation study.}
\label{tab:simulation-factors}
\begin{tabular}{lcc}
\toprule
\textbf{Factor} & \textbf{Low level} & \textbf{High level} \\
\midrule
Treated sample size& $s=500$& $S=2000$\\

Correlation parameter& $c=U(0, 0.1)$& $C=U(0.8, 0.85)$\\

Covariate dimension 
& $p=10$& $P=20$\\

Mixture ratio $(f_1:f_2:f_3)$& $d=4:6:5$& $D=2:8:5$\\
\bottomrule
\end{tabular}
\end{table}

To induce dependency between covariates, the data $\bm{X} \in \mathbb{R}^{N \times P}$ are generated from a lower-dimensional latent variable $\bm{Z} \in \mathbb{R}^{N \times K}$. We generate 20 replicates per scenario, using a hierarchical data-generating process. For more details, see Appendix.~\ref{ap:simulation}.

\subsubsection{Results}

\begin{figure}[H]
    \centering
    \includegraphics[width=\linewidth]{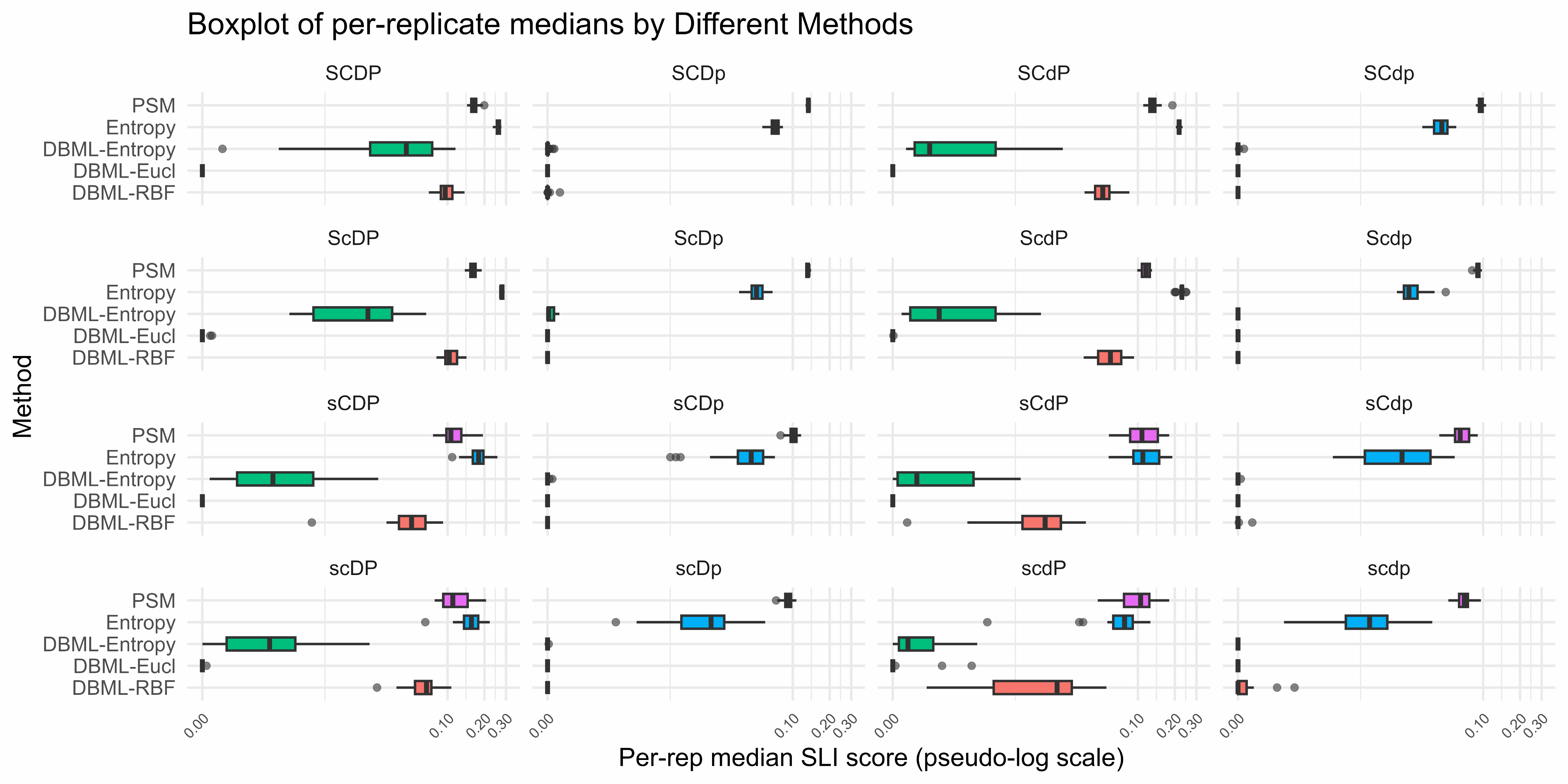}
    \caption{Super Learner Score of different metrics among all simulation design study: Each design contains 20 replicated datasets, and in each replicate we select the final control group 5 times. We take the median SLI for this 5 times selection in each replicates. Design is represented by ``SCDP'' for letters as ``Size'', ``Correlation'', ``Difference'', ``Covariate Dimension'', where the capital letter denotes the higher setting and the lowercase letter indicate the lower setting.}
    \label{fig:boxplot-diff}
\end{figure}

Figure \ref{fig:boxplot-diff} shows that the \gls{method}-based methods generally achieve lower SLI scores than propensity-score matching and standalone entropy balancing across the simulation scenarios. \gls{method}-Eucl provides the best overall performance, with SLI scores consistently close to zero and little variation across replications. Each boxplot summarizes the 20 replication-level median SLI scores, where each replication-level median is calculated over 5 random final-control selections.

Propensity-score matching and standalone entropy balancing show greater imbalance, with entropy balancing deteriorating notably as the number of covariates increases. Although both Entropy and \gls{method}-Entropy use entropy balancing to obtain unit-level weights, \gls{method}-Entropy consistently achieves better covariate balance, particularly in higher-dimensional settings. This improvement highlights the advantage of the \gls{method} pipeline, which removes outliers before weighting. As a result, the pipeline yields more stable balance across replications. 

Among the balancing methods embedded within \gls{method}, \gls{method}-Eucl performs best and remains highly stable across different numbers of covariates. \gls{method}-RBF performs well in lower-dimensional settings but deteriorates as the dimension increases. It is likely because the RBF kernel becomes less discriminative and more sensitive to bandwidth selection in higher dimensions. \gls{method}-Entropy showed a similar pattern, although the decline was generally less severe. It remains competitive, but does not generally achieve the near-zero SLI scores as \gls{method}-Eucl.

More broadly, \gls{method} maintains strong performance across a wide range of simulation settings. It remains effective for mixed covariate types, under varying correlation structures, and when discrete covariates are highly imbalanced. It can identify suitable external controls even when the external and treated distributions differ, provided that the support condition is satisfied.


\begin{figure}[htbp]
    \centering
    
    \begin{subfigure}[b]{0.8\linewidth}
        \centering
        \includegraphics[width=\linewidth]{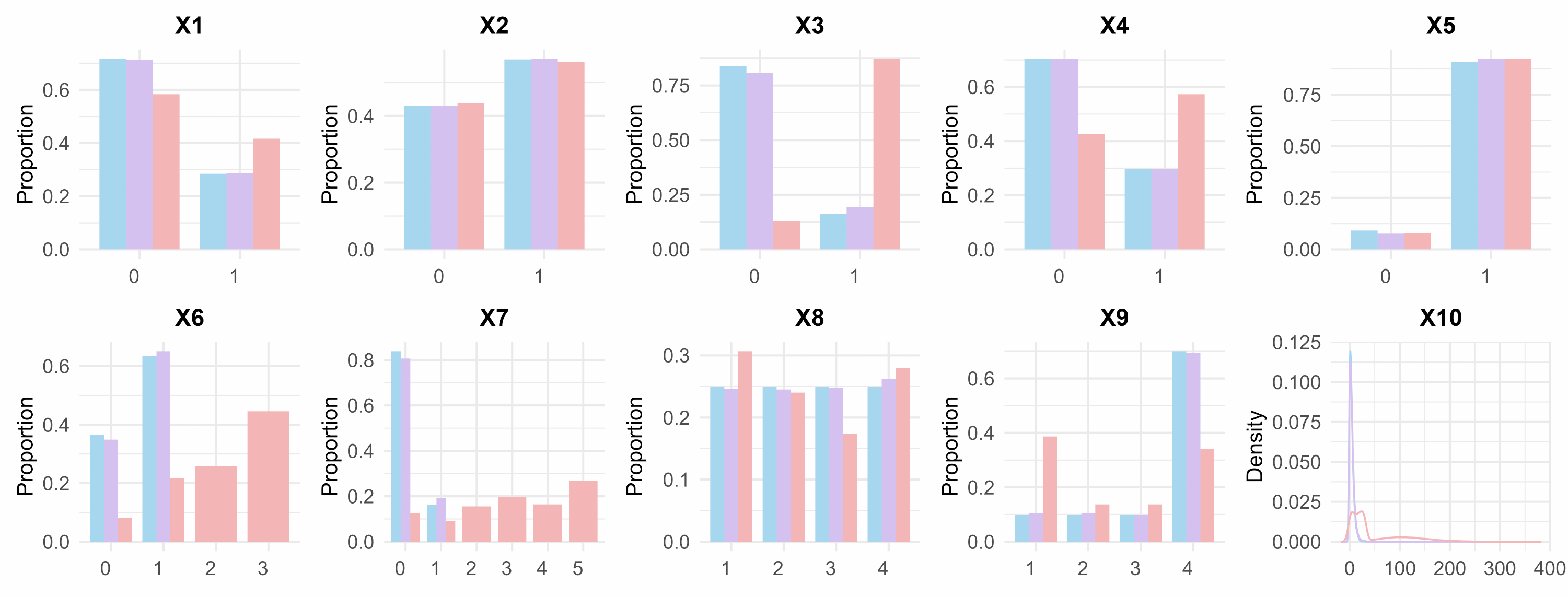}
        \caption{Distribution differences after anomaly detection}
        \label{fig:dist10-1}
    \end{subfigure}
    \vspace{1em}
    \begin{subfigure}[b]{0.8\linewidth}
        \centering
        \includegraphics[width=\linewidth]{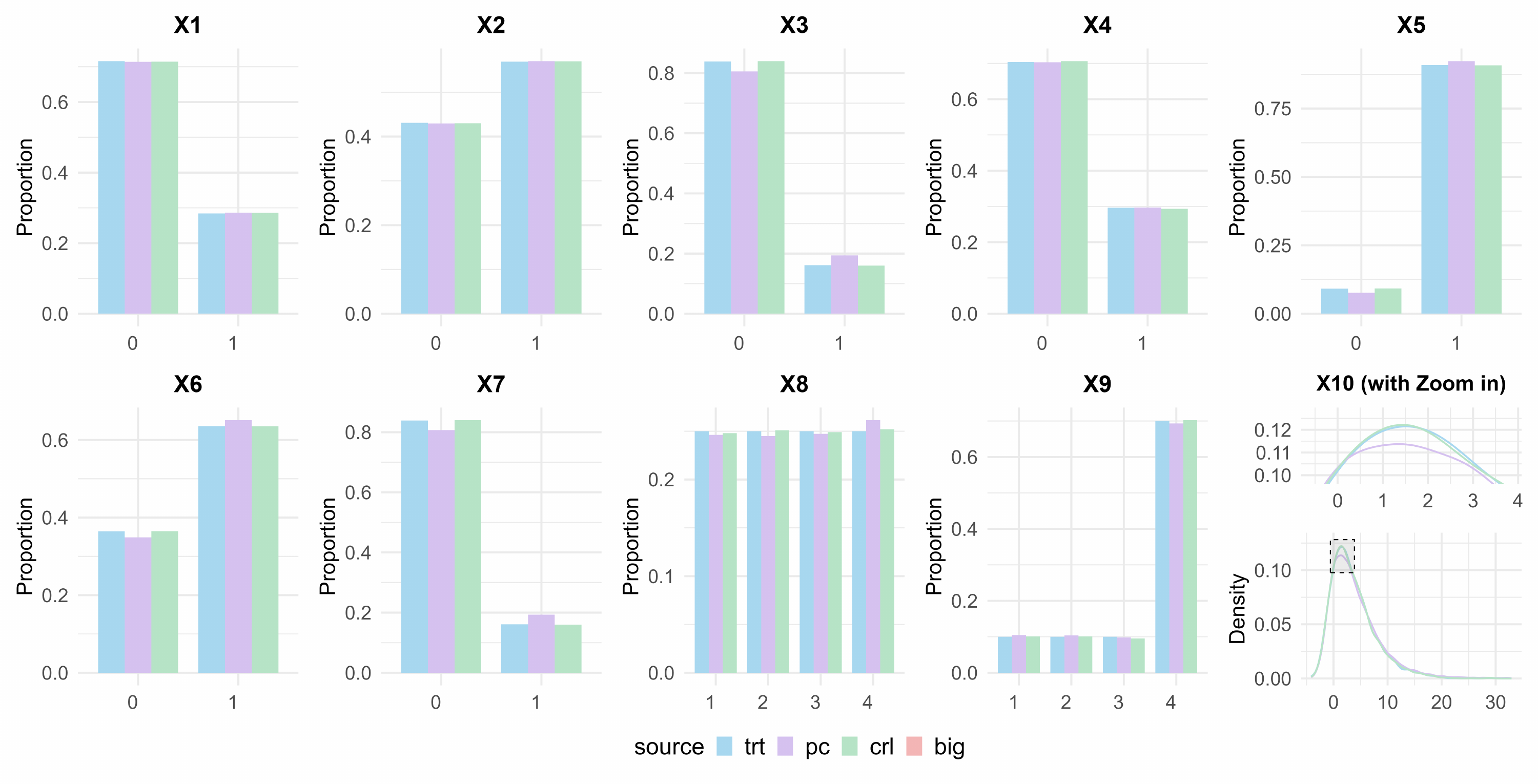}
        \caption{Distribution differences after MMD/LPM}
        \label{fig:dist10-2}
    \end{subfigure} 
    \caption{Distribution plot for each variables for 10-dim case: Comparing big database, possible controls after anomaly detection, the final selected control group (\gls{method}-Eucl) and treatment group.}
    \label{fig:dist10_sub}
\end{figure}

Figures~\ref{fig:dist10_sub} illustrate how the \gls{method} pipeline progressively improves covariate balance. The original external database differs noticeably from the treated group for several discrete covariates and for the continuous covariate $X_{10}$. After anomaly detection, the potential-control group more closely resembles the treated group, indicating that this step effectively removes observations that are poorly supported by the treated covariate distribution. After MMD weighting and LPM selection, the distributions of the final selected controls closely align with those of the treated group across both discrete and continuous covariates.

\begin{figure}[H]
    \centering
    \includegraphics[width=\linewidth]{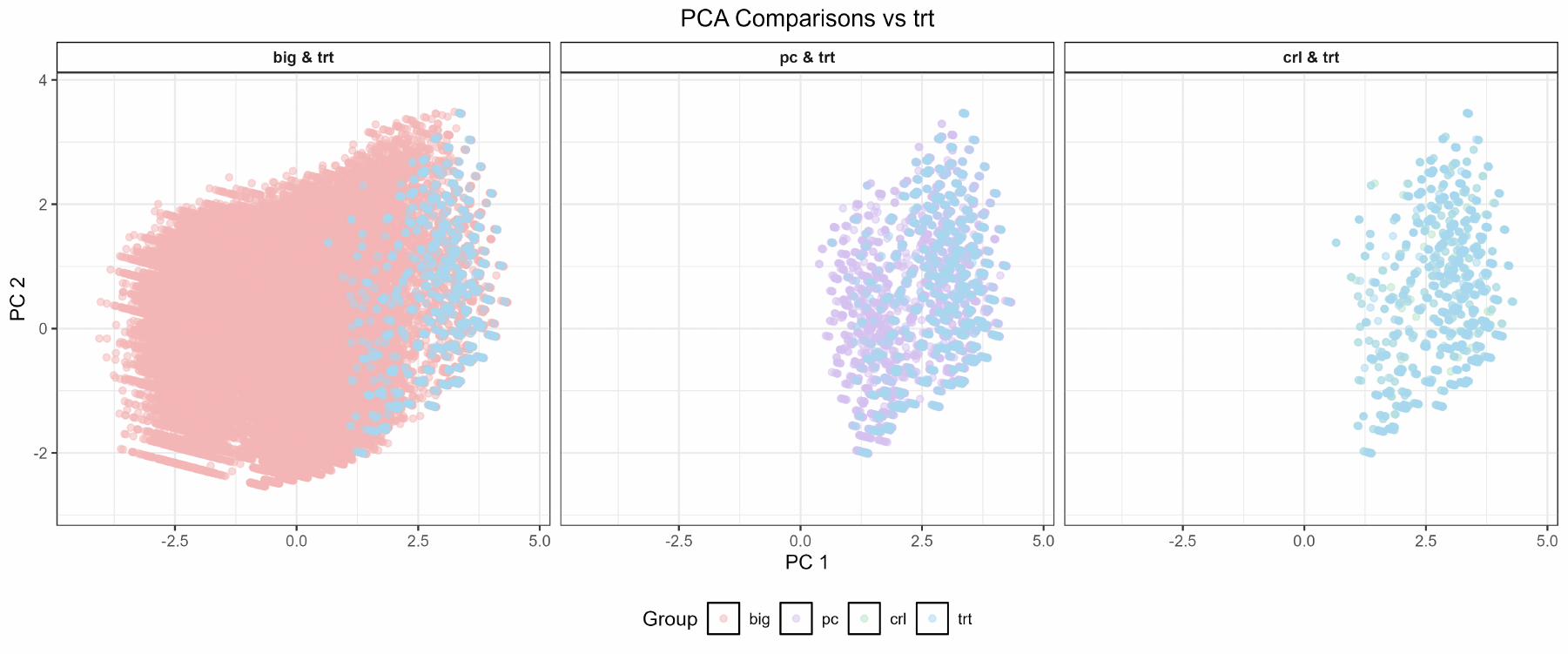}
    \caption{PCA for 10-dim case: Comparing big database, possible controls, final selected control (\gls{method}-Eucl) and treatment group.}
    \label{fig:pca10}
\end{figure}

The PCA plots in Figure~\ref{fig:pca10} provide a view of the joint covariate distribution. The treated group occupies a relatively small region of the broader external-data distribution. After the anomaly-detection step, the possible-control group becomes much more concentrated within the same region as the treated group. The subsequent MMD weighting and LPM selection further improve it, with the final selected controls showing similar location, spread, and overall shape to the treated group in the first two principal-component directions.

Together, these results illustrate the complementary roles of the two stages of \gls{method}. Anomaly detection performs an initial screening to remove unsuitable external controls, while MMD weighting and LPM selection further refine the remaining pool to achieve closer distributional balance with the treated population. This provides visual evidence that the \gls{method} pipeline improves balance progressively, rather than relying on a single matching or weighting step.

\subsection{Synthetic dataset based on PACER}

The Placental Abruption and Cardiovascular Event Risk (PACER) study is a retrospective, population-based birth cohort study in New Jersey, USA \citep{ananth2024pacer}. It links vital statistics, hospital discharge, and mortality records from 1993 to 2020 to track longitudinal maternal and offspring health. The cohort includes 1,877,824 birthing persons and 3,093,241 deliveries ($\geq$ 20 weeks of gestation), with 33,058 cases of placental abruption (1.1\% prevalence). The dataset includes features such as subject demographics, pregnancy complications, labor, infant outcomes, cardiovascular events, and deaths. The primary outcome is placental abruption, with secondary outcomes including gestational age and time to cardiovascular events. The data used here \citep{Cabrera2026AC} consist of two synthetic datasets generated from the original data, preserving the mean and the multivariate correlation structure between features. 
The code to generate the data is given by the R function \verb|CopyData| that belongs to the R package \verb|DNAMR| \citep{CabreraDNAMR2026}.

\begin{figure}[H]
    \centering
    \includegraphics[width=\linewidth]{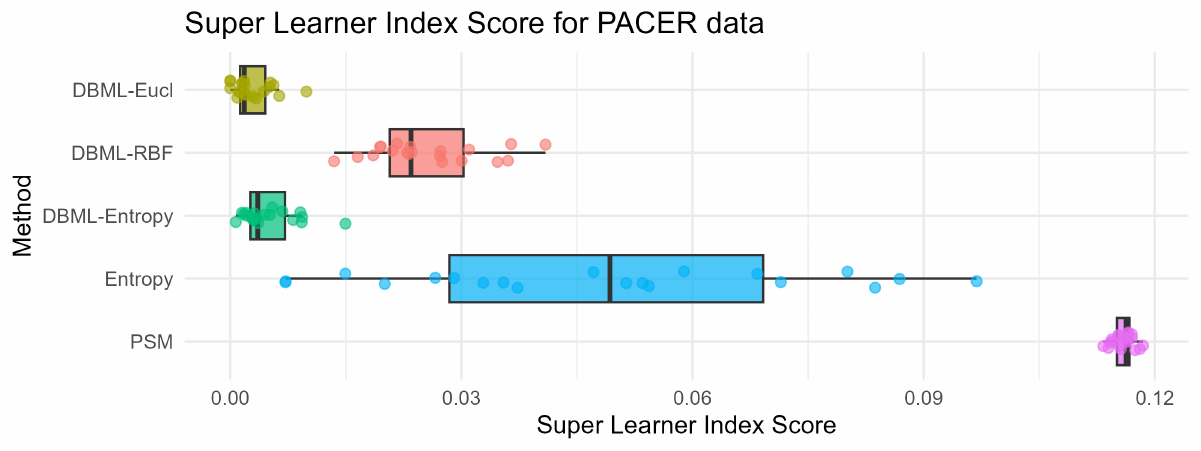}
    \caption{Boxplot for super learner score comparing different metrics: The scatter represents the actual value, each metrics select the final controls 20 times.}
    \label{fig:pacer_sl}
\end{figure}

Figure.~\ref{fig:pacer_sl} shows that the \gls{method}-based methods achieved better covariate balance than propensity-score matching and standalone entropy balancing in the PACER data. Among all methods, \gls{method}-Eucl obtained the lowest SLI scores and showed very little variation across the 20 repeated selections, indicating both strong balance and stable performance. \gls{method}-Entropy also performed well, with SLI scores close to those of \gls{method}-Eucl.

In contrast, standalone entropy balancing showed much larger variability across repeated selections, while propensity-score matching produced the highest SLI scores across runs. The comparison between Entropy and \gls{method}-Entropy again highlights the value of the full \gls{method} pipeline.

The variable-level distributions reported in Figure.~\ref{fig:pacer_sub}, \ref{fig:pacerpca}, \ref{fig:pacerumap} of the Appendix additionally show that the \gls{method} pipeline improves the distributional balance progressively: anomaly detection improves the initial comparability of the possible controls, and the subsequent MMD/LPM further aligning the final controls with the treated group.

\subsection{Real-world dataset: MIMIC-IV}
We constructed a dataset using MIMIC-IV and the MIMIC-IV-Ext-MDS-ED benchmark dataset \citep{PhysioNet-mimiciv-3.1, johnson2023mimiciv, PhysioNet-multimodal-emergency-benchmark-1.0.0, goldberger2000physiobank}. MIMIC-IV is a large public electronic health record database, and MDS-ED is an emergency-department benchmark derived from MIMIC-IV resources. After data integration and preprocessing, the study population included 45,422 patients and 15 covariates. We randomly sampled 1,000 patients with Medicaid insurance as the target sample. $N=37{,}647$ patients with other insurance types formed the external candidate pool. Additional details on this data and preprocessing are provided in Appendix~\ref{app:mimic_preprocessing}.

\begin{figure}[H]
    \centering
    \includegraphics[width=\linewidth]{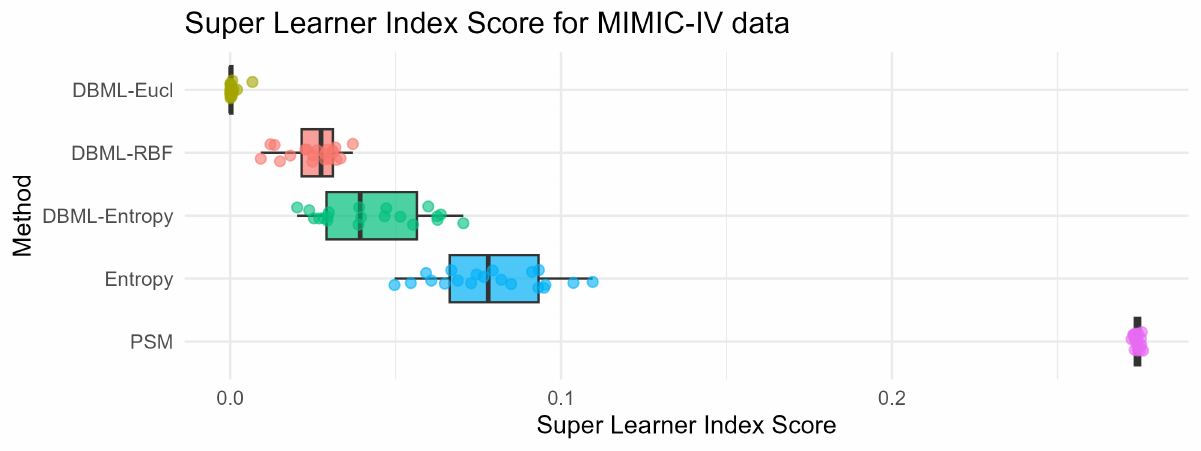}
    \caption{Boxplot for super learner score comparing different metrics: The scatter represents the actual value, each metrics select the final controls 20 times.}
    \label{fig:mimic_sl}
\end{figure}

The overall pattern is consistent with the PACER results, Figure~\ref{fig:mimic_sl} shows that the \gls{method}-based methods achieve better covariate balance than propensity-score matching and standalone entropy balancing in the MIMIC-IV application. \gls{method}-Eucl performs best, with SLI scores concentrated near zero and little variation across repeated selections. However, \gls{method}-RBF outperforms \gls{method}-Entropy in MIMIC-IV, whereas the reverse is observed in PACER. It suggests that their relative performance may depend on the covariate structure of the dataset.

Additional distributional comparisons are provided in Figures~\ref{fig:mimic_sub}, \ref{fig:mimicpca}, and \ref{fig:mimicumap} in the Appendix. Consistent with the PACER results, the pipeline progressively improves distributional balance between the external control and treated groups.

\section{Discussion}
\label{sec:diss}
This paper proposed a distributional balancing pipeline for selecting comparable controls from a large real-world observational database. The method integrates VAE-based anomaly detection, MMD-based distributional comparison, and LPM-based balanced sampling to improve covariate balance between treated and selected control groups. Across simulation studies and experiments using real datasets, the proposed procedure achieved better covariate distributional balance than alternative methods. Based on our results of experiments, the \gls{method} with Euclidean kernel in MMD is generally recommended.

As noted in Section~\ref{sec:intro}, the proposed \gls{method} can be extended to settings where the clinical dataset already contains a group of existing controls. In this setting, the anomaly detection step remains unchanged. The MMD step then minimizes the distributional distance between the treated group and the potential combined control group, which consists of the existing controls and the external possible controls. During this step, only the weights of the external possible controls are estimated, while the weights of the existing controls are fixed at 1. Finally, the LPM selection step is applied only to the external possible controls to determine which additional controls are included in the final combined control group.

Our method has some limitations. Firstly, the use of VAE-based anomaly detection requires careful consideration of the cutoff selection, namely the quantile value $q$. In the absence of overfitting, we empirically find that setting $q=1$ performs well. We recommend assessing potential overfitting of the VAE anomaly detector using diagnostics such as cross-validation among treated units and PCA or UMAP visualizations. Moreover, VAE-based anomaly detection inherits the general limitations of deep learning methods, including sensitivity to hyperparameter tuning. In particular, the dimension of the latent variable in each VAE should be selected carefully; practical recommendations are provided in Appendix~\ref{sec:trick}.

Second, our approach inherits the known limitations of MMD in high-dimensional settings, although these limitations may be partially alleviated by using a kernel based on Euclidean distance. In addition, when a large number of potential controls remain for comparison, the memory requirements of the MMD algorithm can become substantial. Therefore, approximation strategies may be needed to reduce memory usage and improve computational scalability \citep{zhao2015fastmmd}.

Third, \gls{method} is only appropriate when the support of the real-world database covers the support of the treated population. This is common to all distributional matching methods. 

Finally, like other covariate balancing methods for observational data, the proposed approach addresses imbalance in observed covariates and does not by itself eliminate bias from unmeasured confounding. Its validity therefore depends on the availability of covariates that adequately capture the factors related to treatment assignment and outcome variation.


\bibliographystyle{agsm-etal}
\bibliography{reference}

\clearpage
\appendix
\section{Variational autoencoder details}\label{app:vae-details}

For data $\bm{x}_i\in\mathbb{R}^D$,    a deep generative model with Gaussian additive noise is:
\begin{align}
\bm{z}_i &\sim \mathcal{N}(\bm{0},\bm{I}_K) \notag\\
\bm{x}_i &= f_{\theta}(\bm{z}_i) + \bm{\varepsilon}_i,\quad \bm{\varepsilon}_i \stackrel{iid}{\sim}
 \mathcal{N}(\bm{0},\bm{I}_G) \label{eq:dgm-gauss}
 \end{align}
where $\bm{z}_i\in\mathbb{R}^K$ is a per-sample latent variable and $f_{\theta}:\mathbb{R}^K \to \mathbb{R}^D$ is a neural network parameterized by $\theta$. 

A standard variational autoencoder (VAE, \citet{kingma2014auto}) fits this deep generative model using amortized variational inference: specifically, the latent variable posterior is approximated with the variational density:
\begin{align}
    q_{\phi}(\bm{z}_i|\bm{x}_i) = \mathcal{N}\big(\mu_{\phi}(\bm{x}_i), \operatorname{diag}(\sigma_{\phi}^{2}(\bm{x}_i))\big),
\end{align}
where $\mu_{\phi}:\mathbb{R}^D\to\mathbb{R}^K$ and $\sigma_{\phi}^{2}:\mathbb{R}^D\to\mathbb{R}^K$ are neural networks.

\begin{figure}[H]
    \centering
    \includegraphics[width=\linewidth]{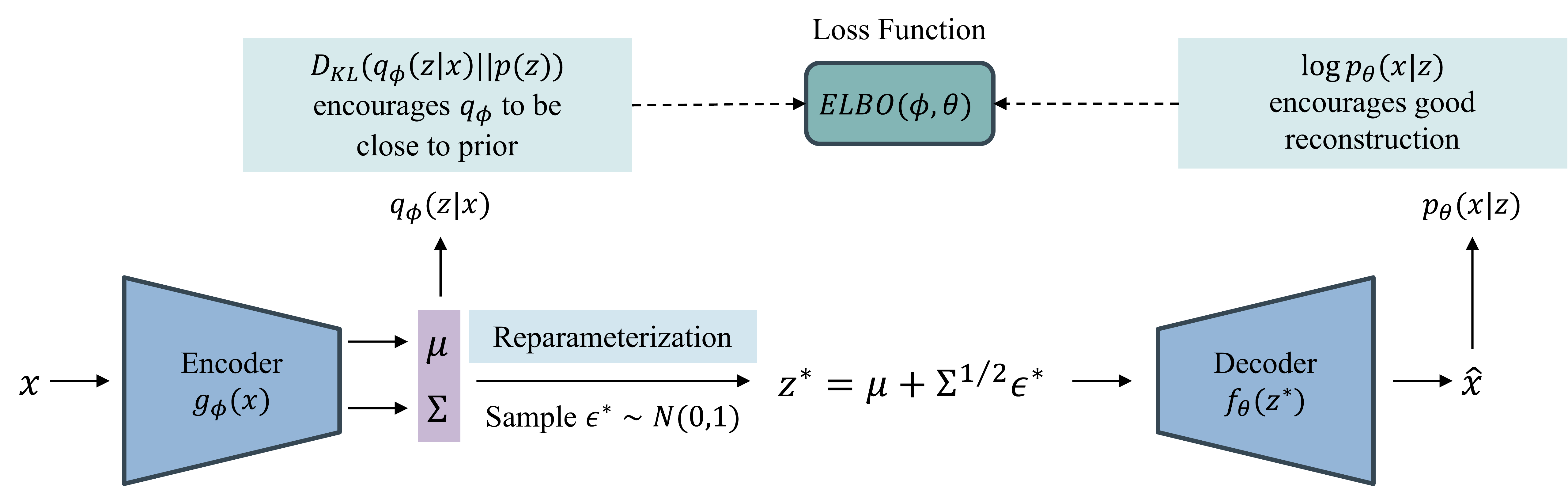}
    \caption{Standard Variational Autoencoder.}
    \label{fig:std vae}
\end{figure}

The VAE is trained by maximizing the evidence lower bound (ELBO) with respect to the encoder and decoder parameters $\{\theta,\phi\}$:
\begin{align}
    \mathrm{ELBO}(\theta,\phi;\bm{X}) = 
    \sum_{i=1}^{N}
   \left\{ \mathbb{E}_{q_{\phi}(\bm{z}_i|\bm{x}_i)}
    \big[\log p_{\theta}(\bm{x}_i|\bm{z}_i)\big]
    -
    D_{\mathrm{KL}}\big(q_{\phi}(\bm{z}_i|\bm{x}_i)\,\|\,p(\bm{z}_i)\big)\right\}.
    \label{eq:elbo}
\end{align}
The first term encourages accurate reconstruction of the observed data, whereas the KL divergence term regularizes the approximate posterior toward the prior distribution. In the ELBO, the first expectation term is approximated via Monte Carlo, combined with the ``reparameterization trick'' (Figure~\ref{fig:std vae}).

For continuous covariates, we use a Gaussian likelihood, as described above (Eq.~\ref{eq:dgm-gauss}). For binary covariates, the likelihood is:
    \begin{align*}
\bm{x}_{i,\mathrm{bin}}|\bm{z}_{i,\mathrm{bin}} \sim \mathrm{Bernoulli}(f_{\theta{\mathrm{bin}}}(\bm{z}_{i,\mathrm{bin}})),
    \end{align*}
    where $f_{\theta_{\mathrm{bin}}}:\mathbb{R}^K \to [0,1]^{D_{\mathrm{bin}}}$ is a neural network. 
    
For categorical covariates we use softmax to convert the logits into probabilities.

For the ordinal covariates, we use an ordered probit model \citep{Wooldridge2010}. Let $D_{\mathrm{ord}}$ denote the number of ordinal covariates, and let $x_{ij,\mathrm{ord}}\in{1,\ldots,M_j}$ denote the $j$th ordinal covariate for observation $i$, where $M_j$ is the number of ordered levels. Each $x_{ij,\mathrm{ord}}$ is treated as a discretization of a continuous latent variable $x_{ij}^*$, such that
\begin{align*}
    x = k \quad \text{if } \gamma_{k-1} < x^* \le \gamma_k, \quad \text{for } k \in \{1, \dots, M\}
\end{align*}
where$ \gamma_0 = -\infty$, $\gamma_M = \infty$ and  $\bm{\gamma}_j=\{\gamma_{j,1},\dots, \gamma_{j,M_j-1}\}$ are threshold parameters.

The ordered probit model assumes $x_{ij}^* \sim N(f_{\theta_{\mathrm{ord}}, j}(\bm{z}_i), 1)$, where $f_{\theta_{\mathrm{ord}}, j}(\bm{z}_i)$ denotes the $j$th output of the decoder network, $\bm z_i$ is the latent representation for observation $i$, and $\theta_{\mathrm{ord}}$ denotes the parameters of the decoder network for the ordinal covariates. Consequently, the  log-likelihood is:
\begin{align*} 
\mathcal{L}(\bm{\gamma}, \theta_{\mathrm{ord}}|\bm{x}_{i,\mathrm{ord}}, \bm{z}_i)=\sum_{j=1}^{D_{\mathrm{ord}}}\sum^{M_j}_{m=1}\mathbb{I}[x_{ij,\mathrm{ord}}=m]\log [\Phi(\gamma_{j,m}-f_{\theta_{\mathrm{ord}}}(\bm{z}_i)) - \Phi(\gamma_{j,m-1} - f_{\theta_{\mathrm{ord}}}(\bm{z}_i))].
\end{align*}
where $\Phi(\cdot)$ denotes the cumulative distribution function of the standard normal distribution.

\section{MMD discussion}\label{sec:mmd-discussion}

The weighted MMD U-statistic is:
\begin{align}
    \text{MMD}^2_u(\mathcal{F}, X, Y, \omega) = &\frac{1}{m(m-1)} \sum ^m_{i=1} \sum^m_{j \neq i} k(X_i, X_j)  + \frac{1}{n(n-1)} \sum^n_{i=1} \sum^n_{j \neq i} k(Y_i, Y_j)\omega_i \omega_j \notag \\
    &- \frac{2}{mn} \sum^m_{i=1} \sum^n_{j=1}k(X_i, Y_j)\omega_j
\label{eq:mmd-ustat}
\end{align}

When the weights are treated as optimization variables, minimizing the weighted U-statistic objective may create an undesirable optimization bias. Specifically, for kernels with nonzero diagonal entries, such as the RBF kernel, the diagonal terms omitted by the U-statistic depend directly on the weights. Consequently, the weighted U-statistic differs from the corresponding V-statistic by a weight-dependent diagonal correction term, which can be approximately expressed as:

\begin{align}
    \text{MMD}^2_u(\mathcal{F}, X, Y, \omega) =\text{MMD}^2_b(\mathcal{F}, X, Y, \omega) - ||\omega||^2_2
\end{align}

As a result, optimizing the U-statistic objective implicitly encourages sparsity in the weights, which is not part of the intended discrepancy measure. Based on this consideration and our empirical results, we adopt the weighted V-statistic formulation in our implementation. 

\section{LPM processing}
\label{ap:lpm-processing}
The details of the LPM algorithm is in Algorithm~\ref{alg:lpm}:

\begin{algorithm}[htbp]
\caption{Local Pivotal Method (LPM)}
\label{alg:lpm}
\small
\KwIn{Finite population of $N$ units with initial inclusion probabilities $\pi_i$; distance function $d$ for nearest neighbors.}
\KwOut{Final inclusion indicators $\pi'_i \in \{0, 1\}$ for each unit.}

Initialize updated probabilities: $\pi'_i \leftarrow \pi_i$ for all $i$ \;

\While{there exist unfinished units ($\pi'_i \notin \{0,1\}$)}{
    Randomly select unfinished unit $i$\;
    
    Find $i$'s nearest neighbor $j = \argmin_{l\in N\backslash i} d(\bm{x}_i, \bm{x}_l)$\;
    \eIf{$j$'s nearest neighbor is $i$}{
        \eIf{$\pi'_i + \pi'_j < 1$}{
            Update $(\pi'_i, \pi'_j)$ as:
            \[
            (\pi'_i, \pi'_j) = 
            \begin{cases}
            (0, \pi'_i + \pi'_j) & \text{w.p. } \frac{\pi'_j}{\pi'_i + \pi'_j}, \\[6pt]
            (\pi'_i + \pi'_j, 0) & \text{w.p. } \frac{\pi'_i}{\pi'_i + \pi'_j}.
            \end{cases}
            \]
        }{
            Update $(\pi'_i, \pi'_j)$ as:
            \[
            (\pi'_i, \pi'_j) = 
            \begin{cases}
            (1, \pi'_i + \pi'_j - 1) & \text{w.p. } \frac{1 - \pi'_j}{2 - \pi'_i - \pi'_j}, \\[6pt]
            (\pi'_i + \pi'_j - 1, 1) & \text{w.p. } \frac{1 - \pi'_i}{2 - \pi'_i - \pi'_j}.
            \end{cases}
            \]
        }
    }{
        Continue
    }
}
\end{algorithm}

The normalized inclusion probabilities $\pi_i$ are not necessarily bounded above by one.  Since LPM requires $0 \leq \pi_i \leq 1$, values larger than one cannot be directly interpreted as inclusion probabilities. In practice, LPM automatically treats units with $\pi_i>1$ as certainty units by setting their inclusion probabilities to $\tilde{\pi}_i=1$. As a result, the total inclusion probability after truncation may be smaller than the target treated sample size $N_{\mathrm{trt}}$, and the final selected control subset may contain fewer than $N_{\mathrm{trt}}$ units.

In the post-adjustment approach, we start from the optimized weights $\hat{\omega}$ and iteratively convert units with adjusted inclusion probabilities larger than one into certainty units. Let $\mathcal{C}$ denote the set of certainty units and let $\mathcal{R}$ denote the remaining possible controls. We initialize $\mathcal{C}=\emptyset$ and let $\mathcal{R}$ be the full set of possible controls. At each iteration, we compute the remaining target sample size as 
\begin{equation} 
n_{\mathrm{remain}} = N_{\mathrm{trt}} - |\mathcal{C}|. \end{equation}
We then rescale the weights of the remaining units by \begin{equation}
\tilde{\pi}_i = \frac{n_{\mathrm{remain}}\hat{\omega}_i} {\sum_{j\in\mathcal{R}} \hat{\omega}_j}, \qquad i\in\mathcal{R}. 
\end{equation} 
If all rescaled probabilities satisfy $\tilde{\pi}_i \leq 1$ for $i\in\mathcal{R}$, the procedure stops. Otherwise, any unit with $\tilde{\pi}_i>1$ is added to the certainty set $\mathcal{C}$, its adjusted inclusion probability is fixed at $\tilde{\pi}_i=1$, and the remaining set $\mathcal{R}$ is updated accordingly. 

This procedure is repeated until all units have valid inclusion probabilities no larger than one. The final adjusted inclusion probabilities are therefore 
\begin{equation} \tilde{\pi}_i = \begin{cases} 1, & i\in\mathcal{C}, \\[6pt] \dfrac{n_{\mathrm{remain}}\hat{\omega}_i} {\sum_{j\in\mathcal{R}}\hat{\omega}_j}, & i\in\mathcal{R}, \end{cases} 
\end{equation} 
where $\mathcal{C}$ and $\mathcal{R}$ are the final certainty and non-certainty sets after convergence. By construction, the adjusted inclusion probabilities satisfy 
\begin{equation} 
0 \leq \tilde{\pi}_i \leq 1 \qquad \text{and} \qquad \sum_i \tilde{\pi}_i = N_{\mathrm{trt}}. 
\end{equation} 
These adjusted probabilities can then be used as valid input inclusion probabilities for LPM. This post-adjustment preserves high-weight units as certainty units while redistributing the remaining sampling mass across the non-certainty units.

\section{Super Learner Index Score}
\label{app:sli}
The Super-Learner Index (SLI) Score\citep{Cabrera2026AC} we used as metric in our paper is:
\begin{align}
\mathrm{SLI}=\mathrm{sd}\left(\{p_i\}_{i=1}^n\right),
\end{align}

Because Super-Learner \citep{vanderlaan2007superlearner} calculates propensity scores using cross-fitting, and thus depends on sampling variability, we calculate the average SLI over multiple splits of the data:
\begin{align}
\overline{\mathrm{SLI}} = \frac{1}{M}\sum_{m=1}^M \mathrm{sd}\left(\{p_i^{(m)}\}_{i=1}^n\right)
\end{align}
where $m$ indexes a split of the data. 
Note that we can replace Super-Learner with any method that calculates the propensity score.

When the treatment and control groups are balanced, the propensity scores should all be close to 0.5 (that is, given the covariates, the probability of being in treatment or control is the same). In this case, the standard deviation of the propensity scores should be minimized; this is what $\overline{\mathrm{SLI}}$ measures.  Consequently, the SLI score is a proxy for dissimilarity of distributions.

To estimate propensity scores, we used the \texttt{SuperLearner} package \citep{SuperLearnerPackage} in R with a binomial outcome model. In the super learner algorithm, we include penalized regression (\texttt{SL.glmnet}), random forests (\texttt{SL.ranger}), multivariate adaptive regression splines (\texttt{SL.earth}), generalized linear models (\texttt{SL.glm}), generalized additive models (\texttt{SL.gam}), and neural networks (\texttt{SL.nnet}). We set  $M=30$ in Eq.~\ref{eq:sli}. 

\section{MIMIC-IV Data Preprocessing}
\label{app:mimic_preprocessing}

The empirical analysis combined information from the MIMIC-IV-Ext-MDS-ED file \verb|mds_ed.csv| and the MIMIC-IV hospital admissions table \verb|admissions.csv|. The MDS-ED data provided demographic characteristics, ethnicity indicators, median heart rate, selected chronic-disease indicators, and an ICU deterioration indicator, while the admissions table provided insurance status, language, and marital status. After preprocessing, the two data sources were merged by patient identifier, resulting in 45,422 patients. 

From \verb|mds_ed.csv|, we selected the patient identifier \verb|general_subject_id|; gender and age; ethnicity indicators for White, Asian, Black/African, and Hispanic/Latino patients; median heart rate; selected ICD-10 diagnosis indicators; and the 24-hour ICU deterioration indicator \verb|deterioration_icu_24h|. The selected ICD-10 diagnosis groups were I10 for essential hypertension, E78 for disorders of lipoprotein metabolism and other lipidemias, E11 for type 2 diabetes mellitus, I25 for chronic ischemic heart disease, K21 for gastro-esophageal reflux disease, and N18 for chronic kidney disease.

The special value \verb|-999| in \verb|deterioration_icu_24h| was recoded as 0. Records with missing values among the selected MDS-ED variables were removed. Because multiple MDS-ED records could correspond to the same patient, one record per patient was retained using \verb|general_subject_id|. After these steps, the MDS-ED-derived dataset contained 56,583 patients and 15 variables.

From the MIMIC-IV \verb|admissions.csv| table, we selected \verb|subject_id|, \verb|insurance|, \verb|language|, and \verb|marital_status|. Missing insurance values were assigned to the category `Other.'' Missing language values were also initially assigned to `Other,'' after which language was grouped into five categories: English, Spanish, Russian, Chinese, and Other. Records with missing marital status were removed. Because a patient could have multiple hospital admissions, one admission record per patient was retained. The resulting admissions-derived dataset contained 213,374 patients and four variables.

The two processed datasets were merged by patient identifier, matching \verb|general_subject_id| in MDS-ED to \verb|subject_id| in the MIMIC-IV admissions table. This merge resulted in 45,422 patients. The merged data were divided according to insurance status. The Medicaid group contained 7,775 patients and was used as the target-domain sample frame. The non-Medicaid group contained 37,647 patients and was used as the external candidate pool. We randomly sampled 1,000 patients from the Medicaid group to construct the target sample, while retaining the full non-Medicaid group as the external candidate pool.

Before applying the proposed \gls{method} method, patient identifiers, insurance status, and \verb|deterioration_icu_24h| were removed. Language and marital status were represented as categorical variables in the implementation. The final analysis therefore used 15 demographic and clinical covariates.

\section{More simulation study result visualization}
\begin{figure}[H]
    \centering
    \includegraphics[width=\linewidth]{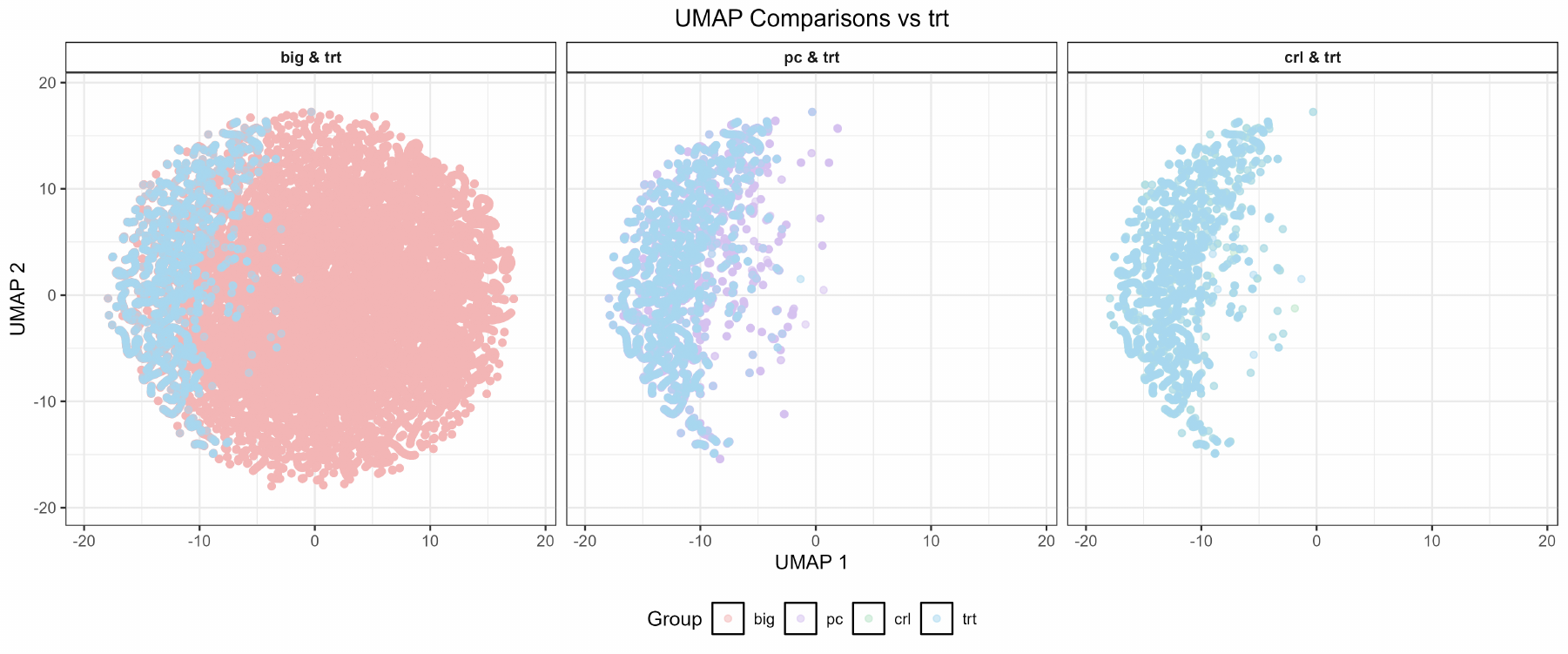}
    \caption{UMAP for 10-dim case: Comparing big database, possible controls, final selected control and treatment group.}
    \label{fig:umap10}
\end{figure}
Dimensionality reduction and visualization were performed using Uniform Manifold Approximation and Projection (UMAP) via the \verb|uwot| package in R. The scaled feature matrix was projected into a two-dimensional embedding space using the Euclidean distance metric. To effectively balance the preservation of local and global geometric structures, the local neighborhood size was set to 15. Additionally, the minimum distance between points in the low-dimensional space was set to 0.1 to optimize cluster resolution and visual clarity. The complete layout model was retained for downstream reproducibility and potential projection of novel data. 

\begin{figure}[H]
    \centering
    \includegraphics[width=\linewidth]{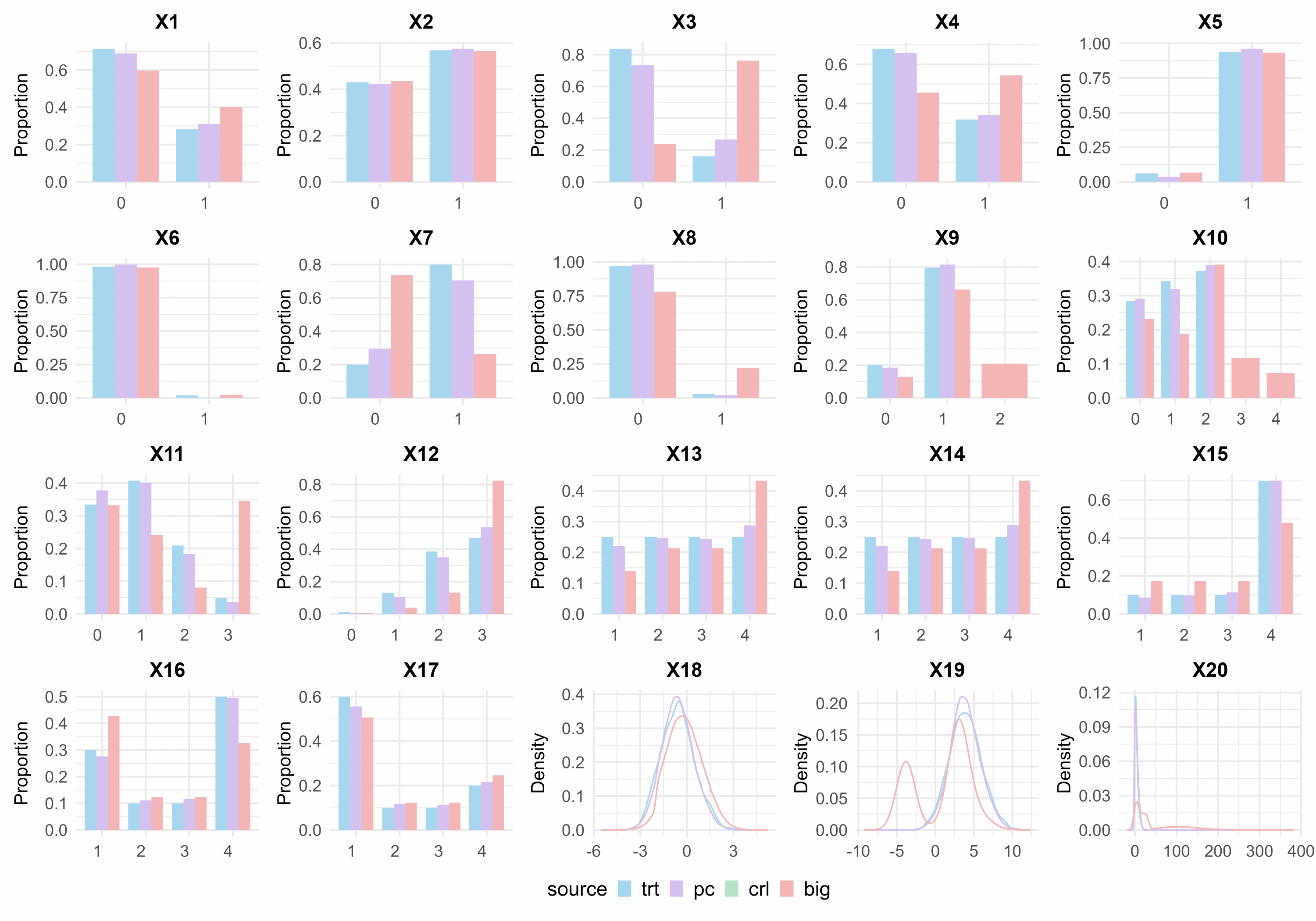}
    \caption{Distribution plot for each variables for 20-dim case: Comparing big database, possible controls after anomaly detection and treatment group.}
    \label{fig:dist20-1}
\end{figure}

\begin{figure}[H]
    \centering
    \includegraphics[width=\linewidth]{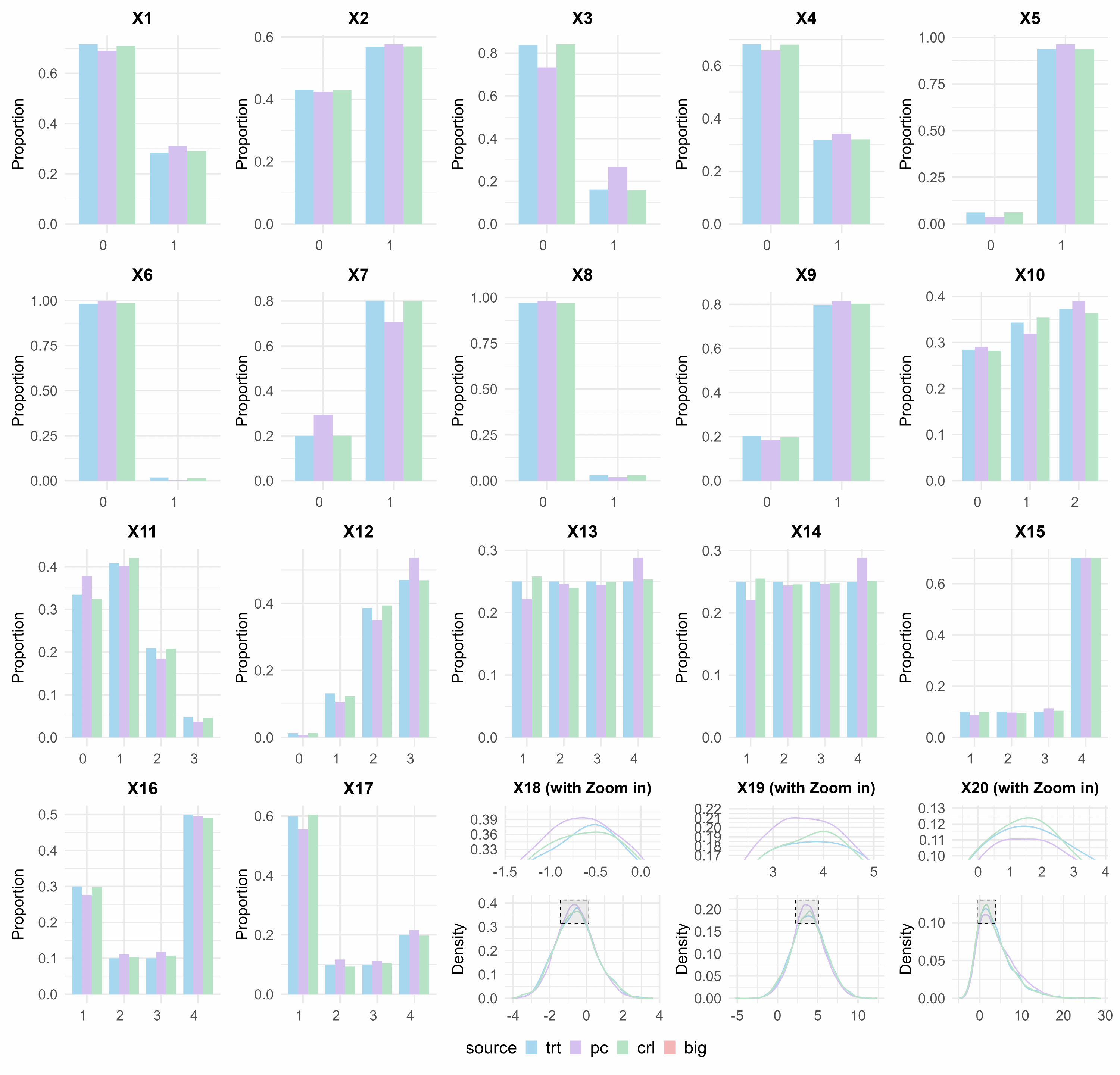}
    \caption{Distribution plot for each variables for 20-dim case: Comparing possible controls after anomaly detection, the final selected control group and treatment group.}
    \label{fig:dist20-2}
\end{figure}

\begin{figure}[htbp]
    \centering
    \includegraphics[width=\linewidth]{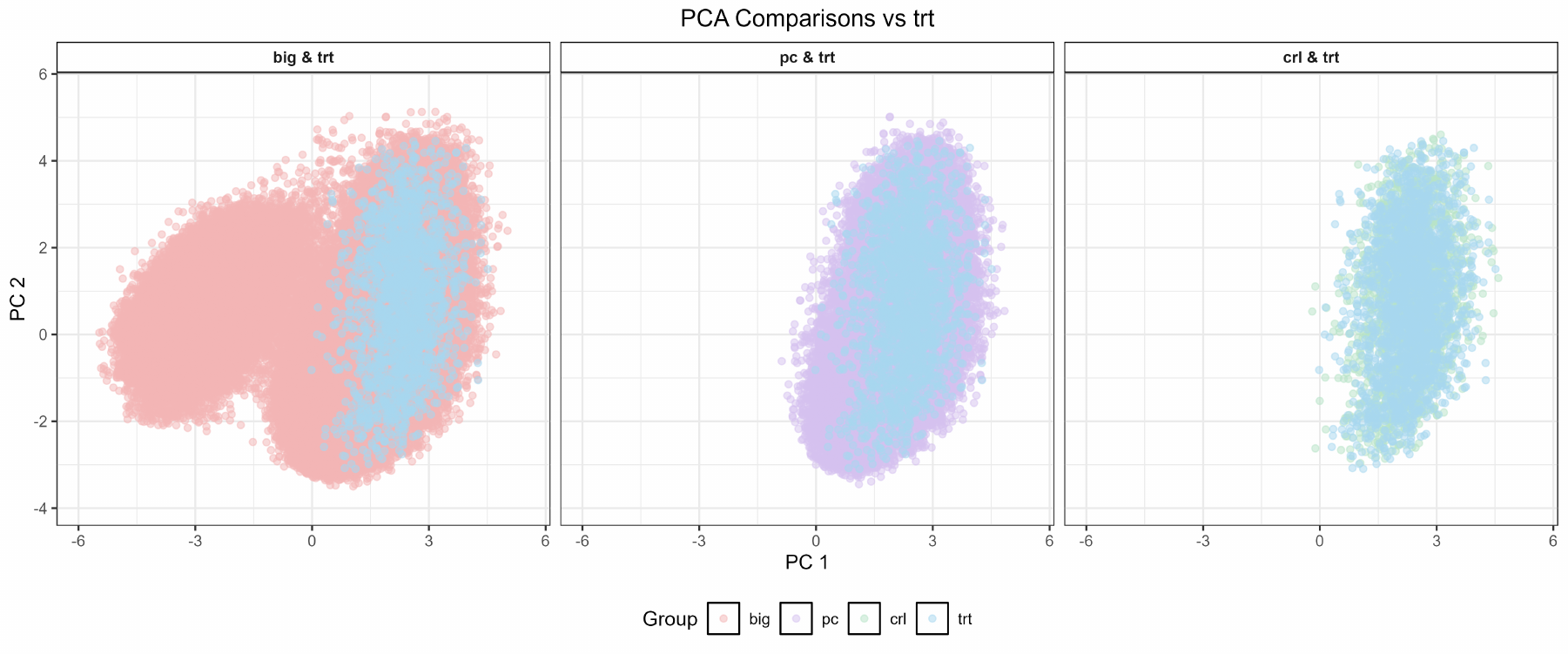}
    \caption{PCA for 20-dim case: Comparing big database, possible controls, final selected control and treatment group.}
    \label{fig:pca20}
\end{figure}

\begin{figure}[htbp]
    \centering
    \includegraphics[width=\linewidth]{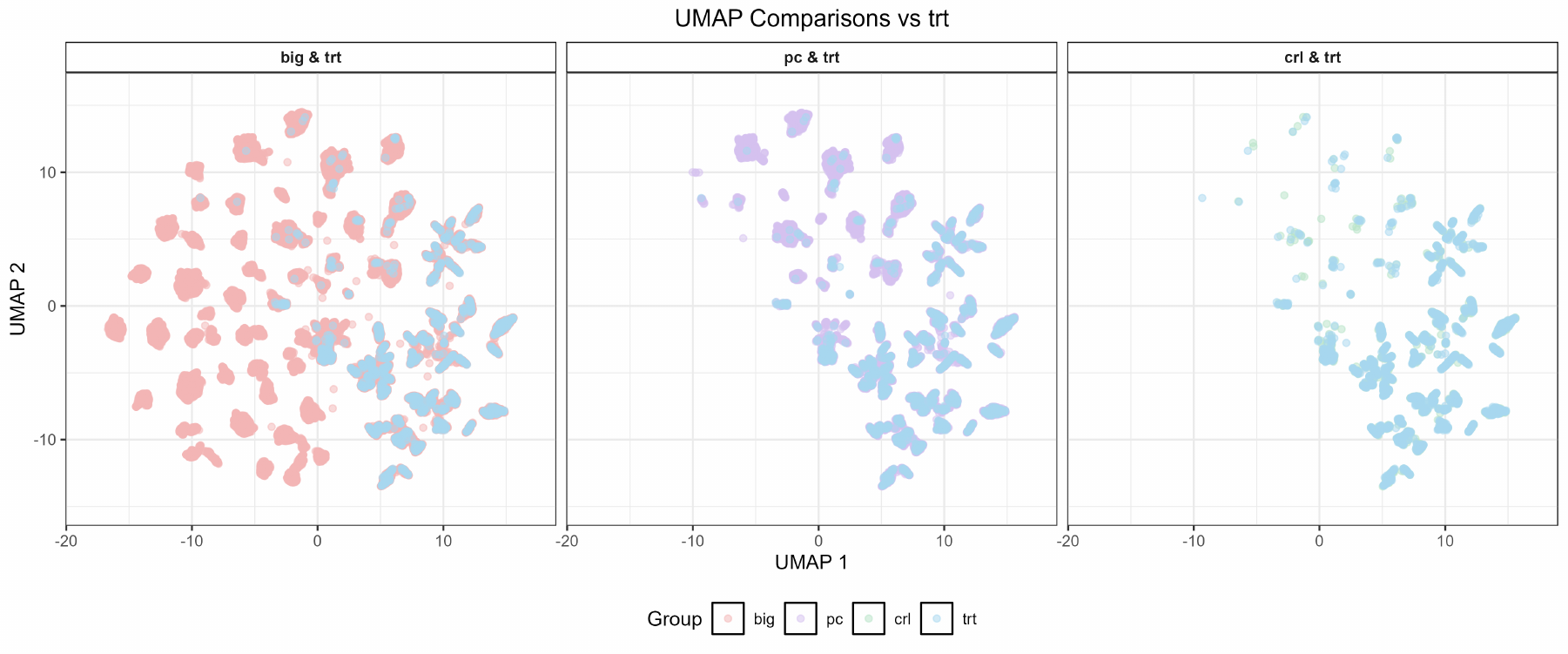}
    \caption{UMAP for 20-dim case: Comparing big database, possible controls, final selected control and treatment group.}
    \label{fig:umap20}
\end{figure}

\begin{figure}[htbp]
    \centering
    \begin{subfigure}[b]{0.7\linewidth}
        \centering
        \includegraphics[width=\linewidth]{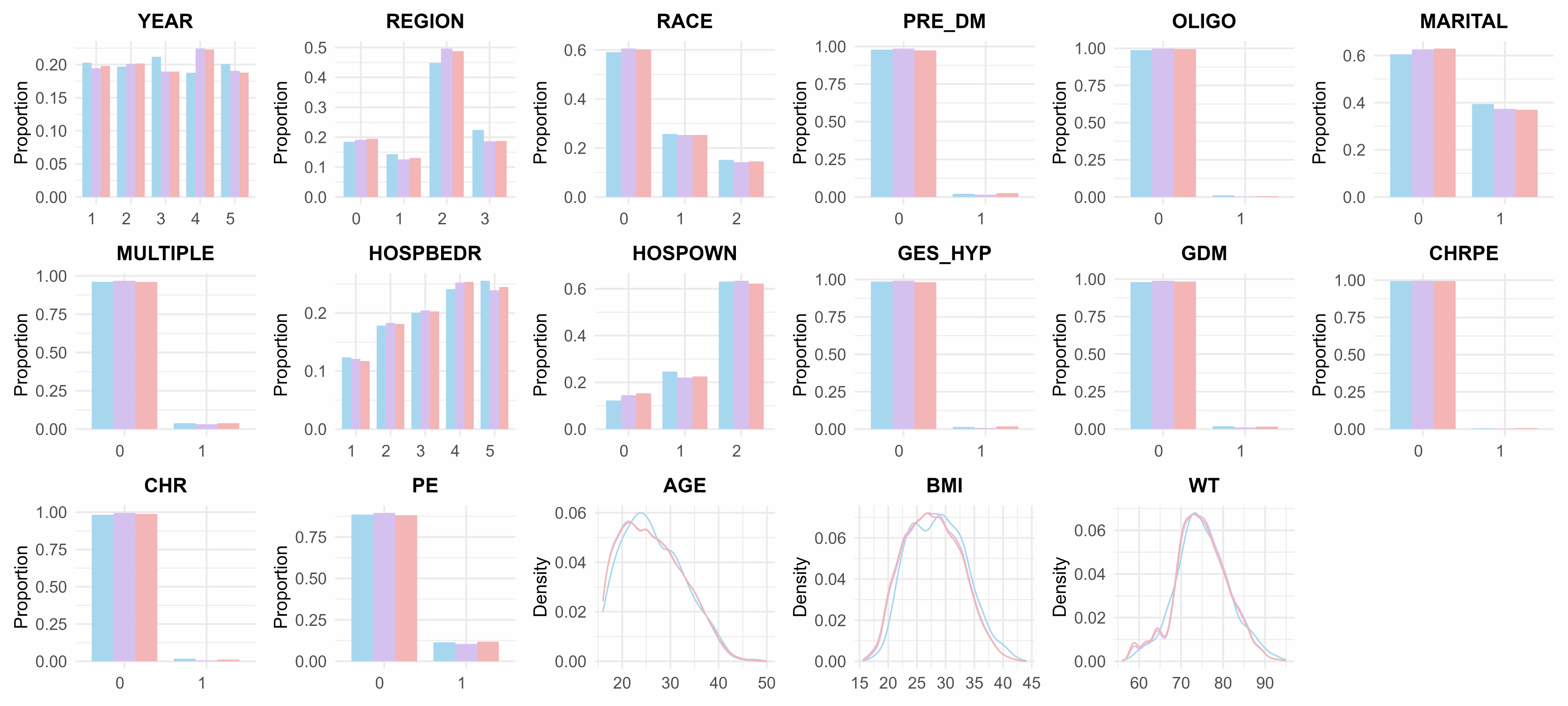}
        \caption{Distribution differences after anomaly detection}
        \label{fig:pacerdist-1}
    \end{subfigure}
    \vspace{1em}
    \begin{subfigure}[b]{0.7\linewidth}
        \centering
        \includegraphics[width=\linewidth]{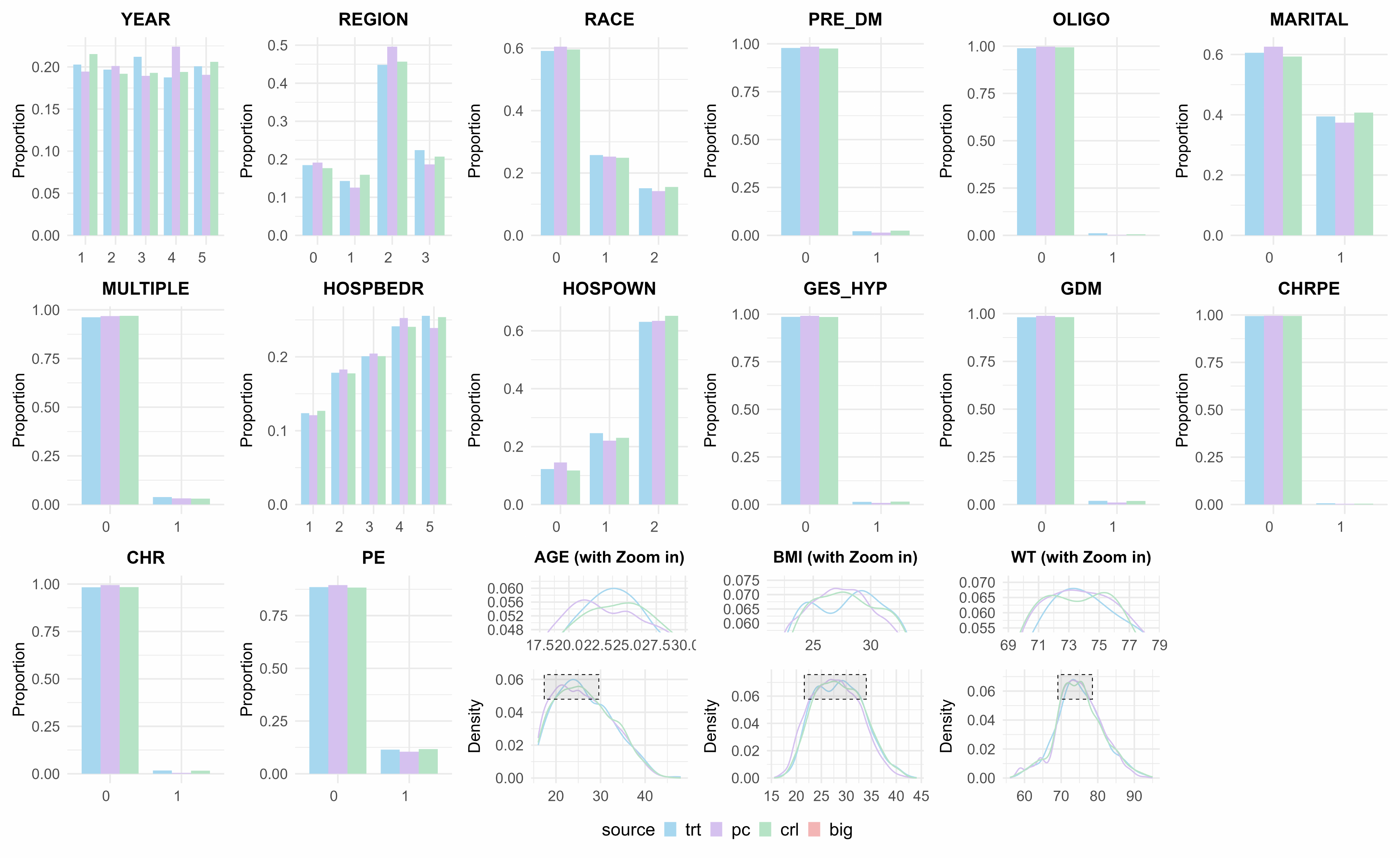}
        \caption{Distribution differences after MMD/LPM}
        \label{fig:pacerdist-2}
    \end{subfigure}
    
    \caption{Distribution plot for each variables for PACER data: Comparing big database, possible controls after anomaly detection, the final selected control group and treatment group.}
    \label{fig:pacer_sub}
\end{figure}

\begin{figure}[htbp]
    \centering
    \includegraphics[width=\linewidth]{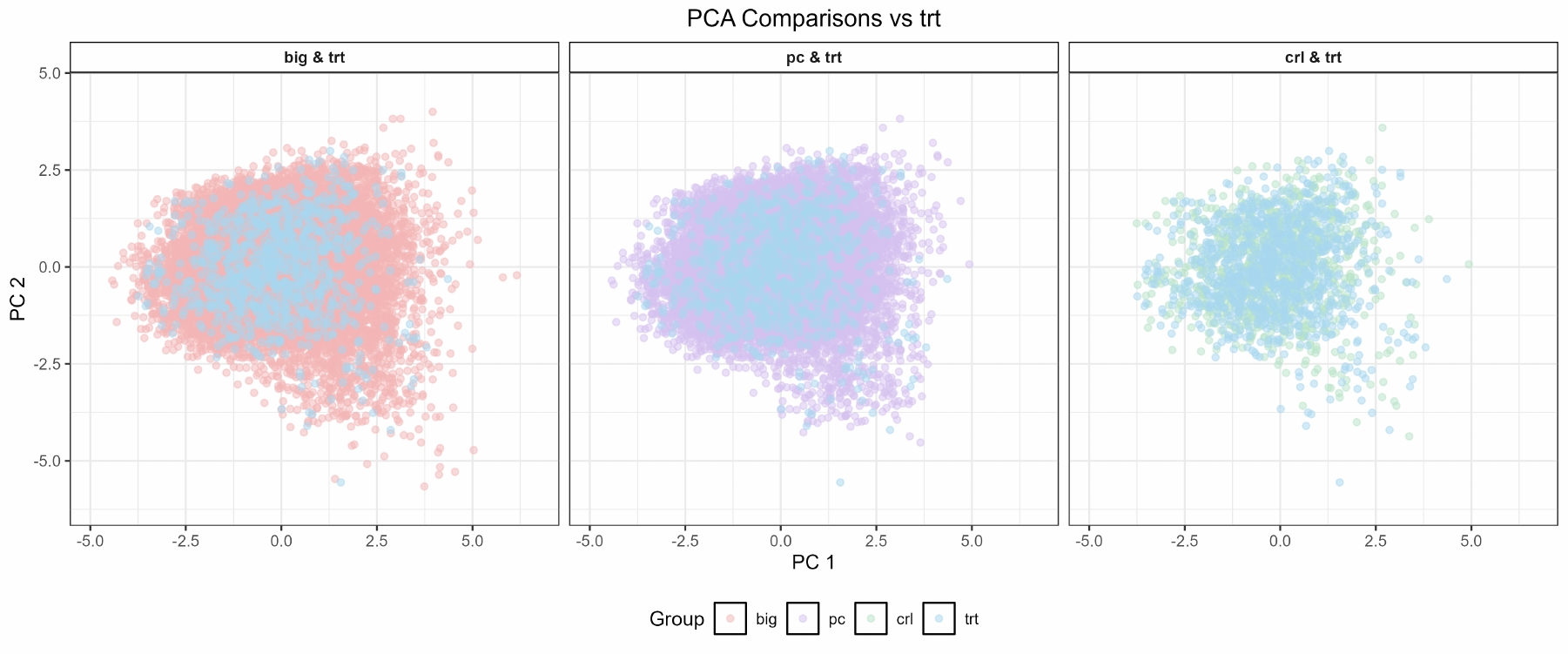}
    \caption{PCA for PACER dataset: Comparing big database, possible controls, final selected control and treatment group.}
    \label{fig:pacerpca}
\end{figure}

\begin{figure}[htbp]
    \centering
    \includegraphics[width=\linewidth]{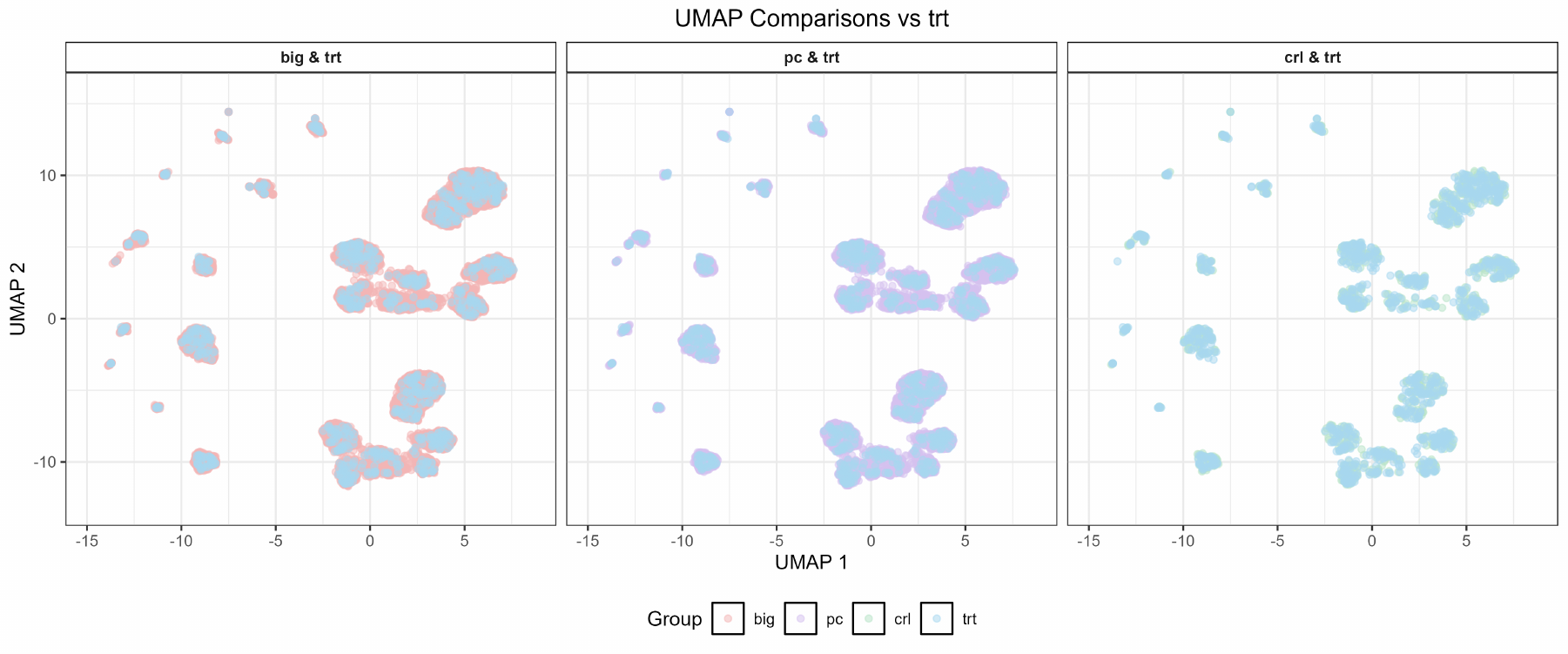}
    \caption{UMAP for PACER dataset: Comparing big database, possible controls, final selected control and treatment group.}
    \label{fig:pacerumap}
\end{figure}

\begin{figure}[htbp]
    \centering
    \begin{subfigure}[b]{\linewidth}
        \centering
        \includegraphics[width=0.7\linewidth]{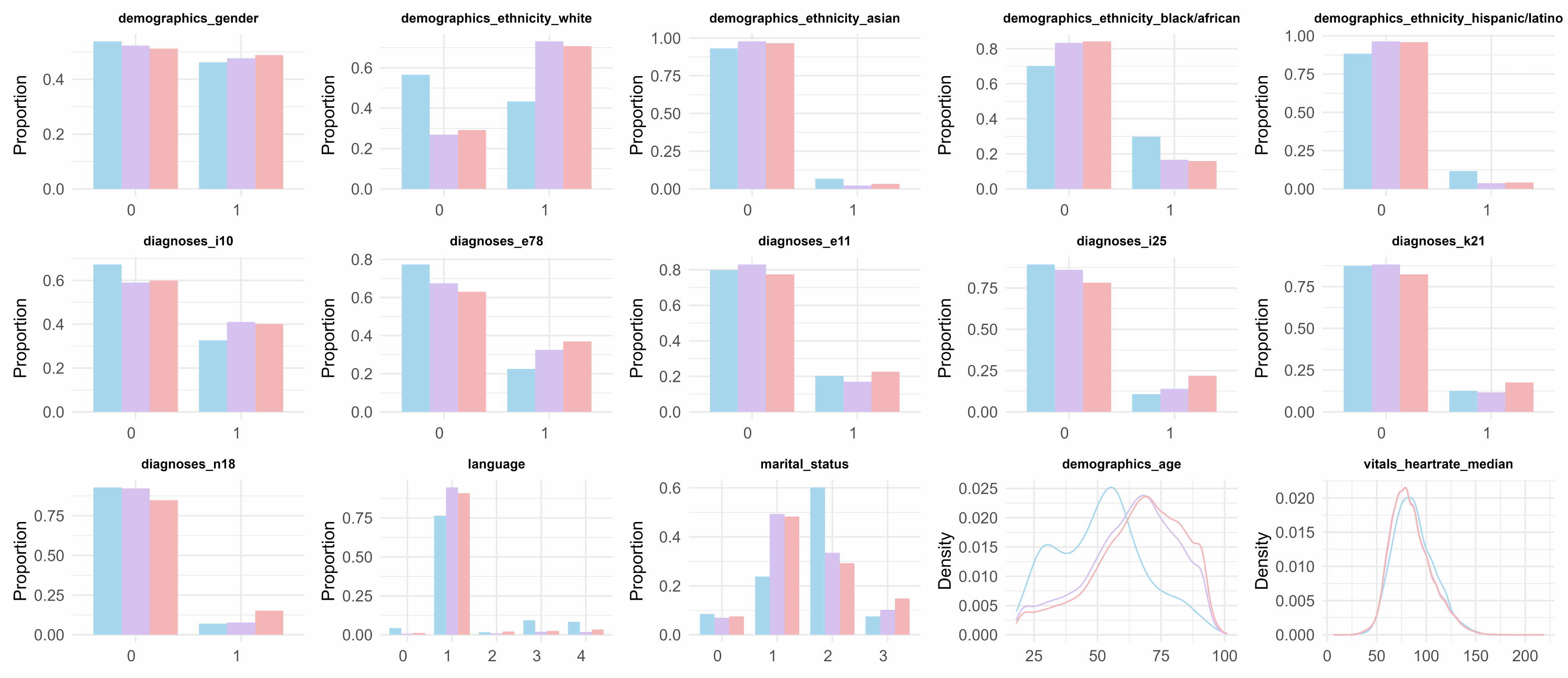}
        \caption{Distribution differences after anomaly detection}
        \label{fig:mimicdist-1}
    \end{subfigure}
    \vspace{1em}
    \begin{subfigure}[b]{\linewidth}
        \centering
        \includegraphics[width=0.7\linewidth]{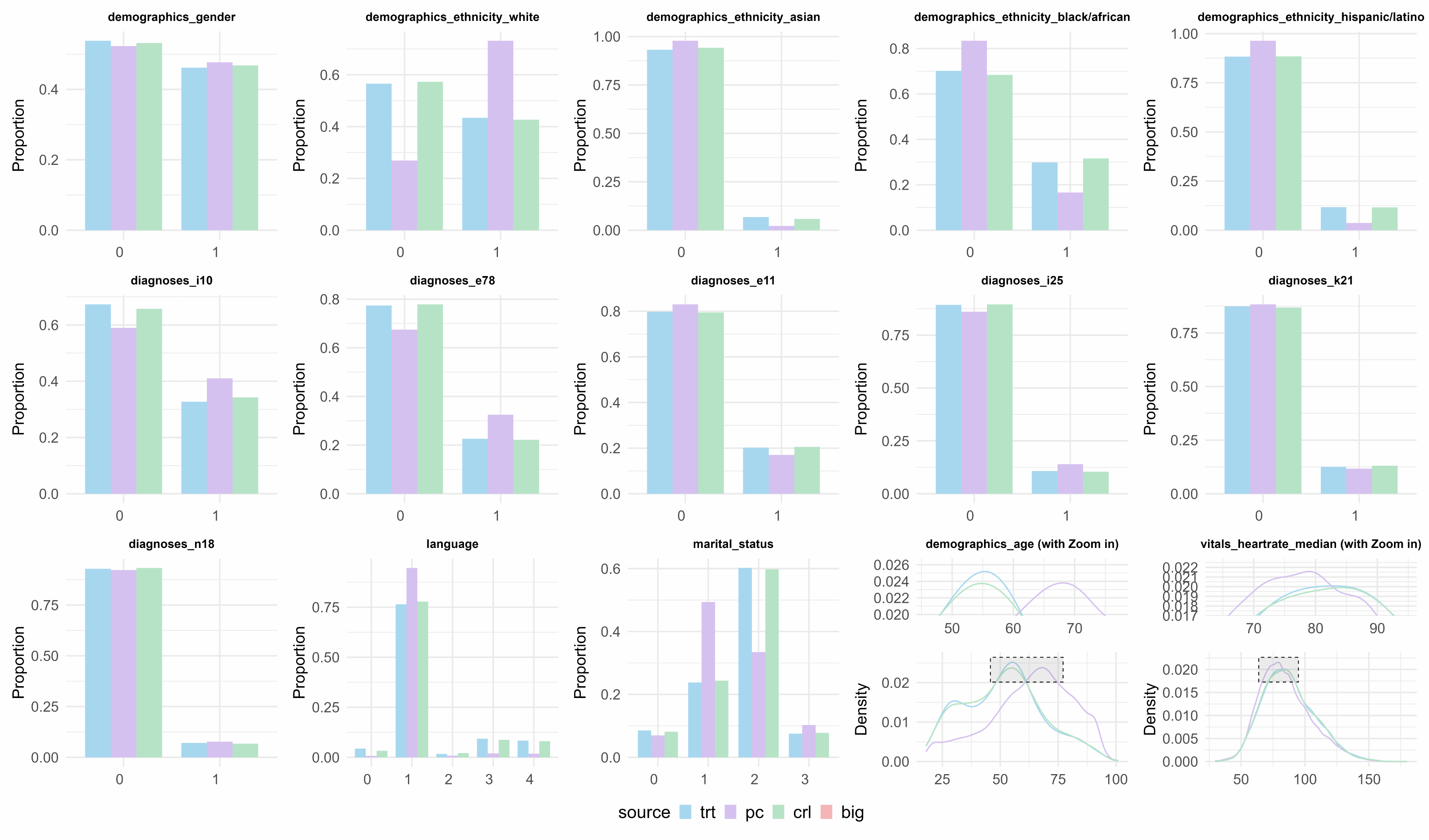}
        \caption{Distribution differences after MMD/LPM}
        \label{fig:mimicdist-2}
    \end{subfigure}
    
    \caption{Distribution plot for each variables for MIMIC-IV data: Comparing big database, possible controls after anomaly detection, the final selected control group and treatment group.}
    \label{fig:mimic_sub}
\end{figure}

\begin{figure}[htbp]
    \centering
    \includegraphics[width=\linewidth]{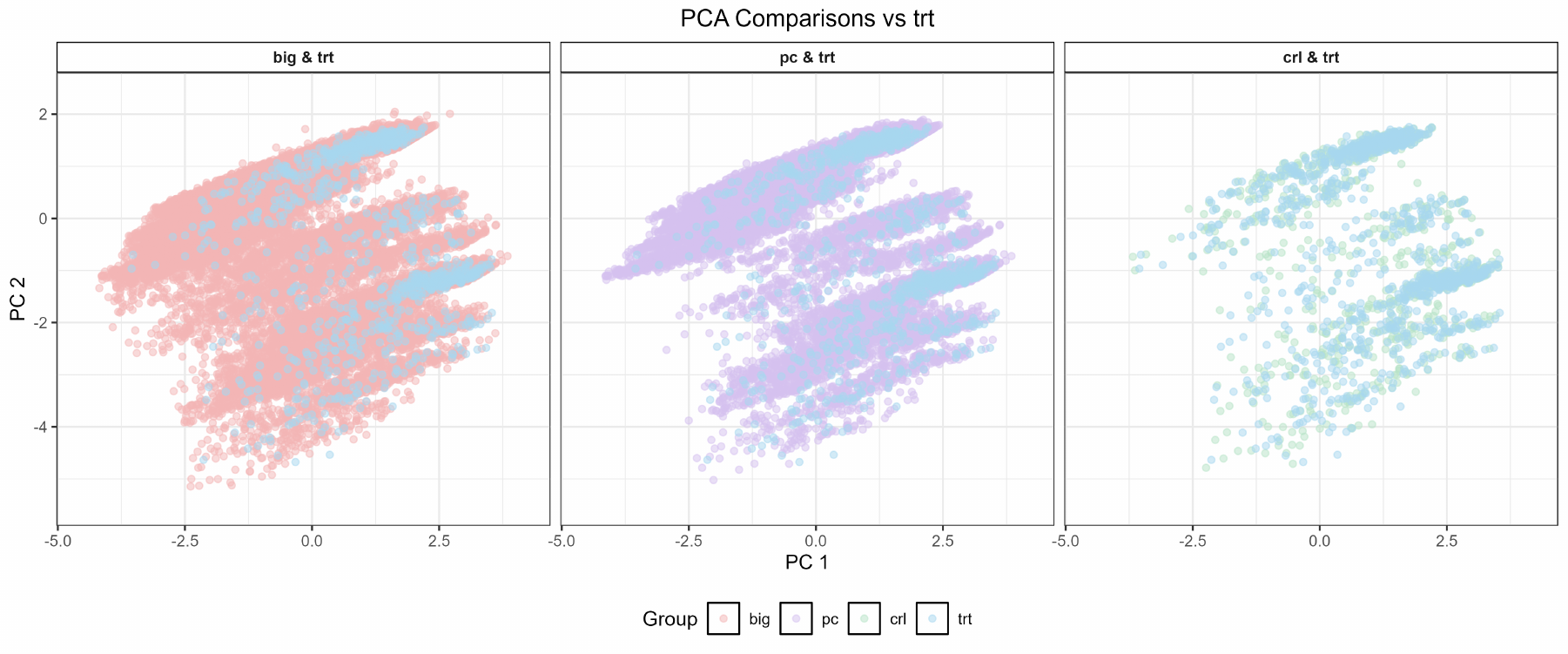}
    \caption{PCA for MIMIC-IV dataset: Comparing big database, possible controls, final selected control and treatment group.}
    \label{fig:mimicpca}
\end{figure}

\begin{figure}[htbp]
    \centering
    \includegraphics[width=\linewidth]{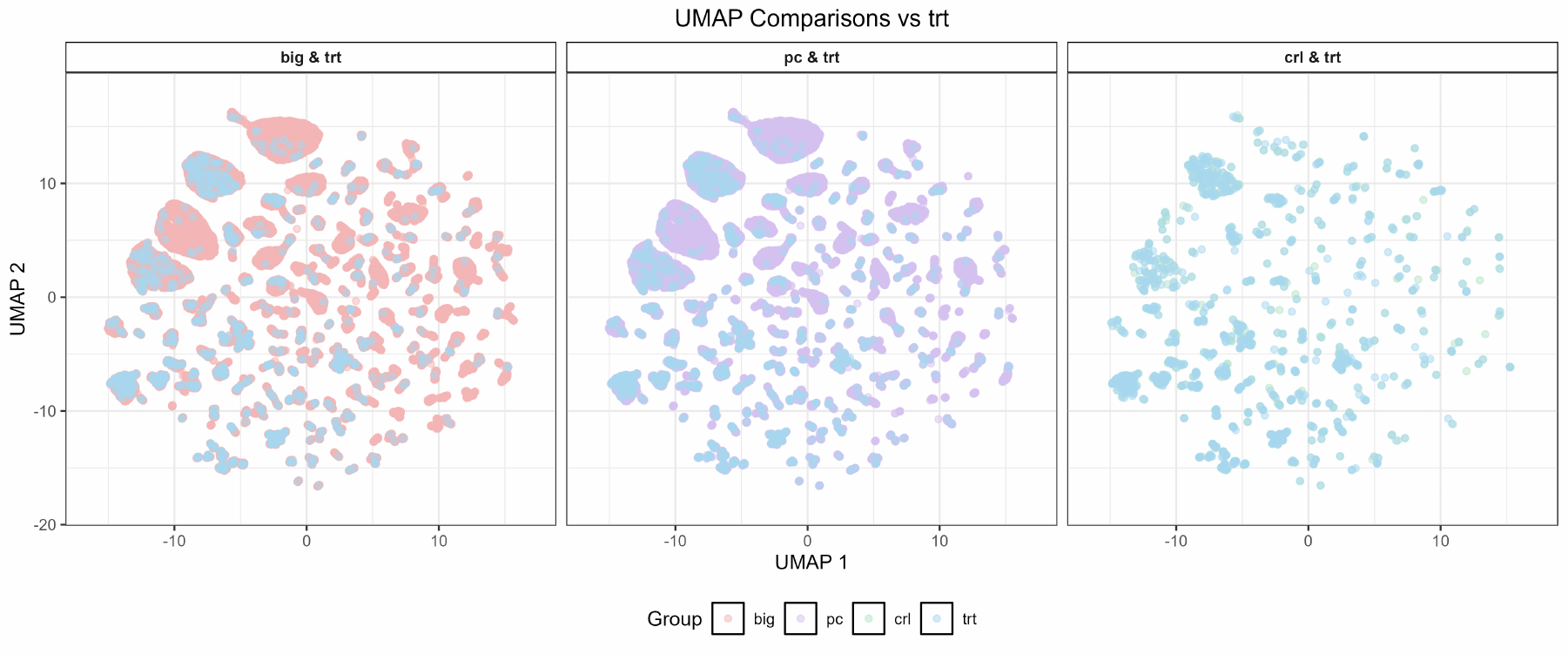}
    \caption{UMAP for MIMIC-IV dataset: Comparing big database, possible controls, final selected control and treatment group.}
    \label{fig:mimicumap}
\end{figure}

\section{75-dim Case}
We also implement our method on extreme case with high dimensional data: 50 dimension of binary variables, 5 dimension of categorical variables, 10 ordinal variables, and 10 continuous variables simulated from the following structure:

\textbf{The distribution setting for $f_1$}
\begin{itemize}
   \item $Z \sim N(\bm{\mu} = (1, 2, 3)^T, \bm{\Sigma} = (d=1, \rho=0.8))$.
    \item Binary Variables:\\
    For $j = 1, \dots,50$,\\
    sample $z_j \sim \mathrm{Unif}\{Z_1, Z_2, Z_3\}$, $\beta_0 \sim \mathrm{Unif}(-1.5, 1.5)$, $\beta_1 \sim \mathrm{Unif}(0.2, 1.0)$;\\
    $X_j \sim Binomial(n=1, p=\sigma(\beta_0 +\beta_1z_j))$.

    \item Categorical Variables:\\
    For $j=1, \dots, 5$,\\
    set the categorical level to be $C = \{3, 4, 5, 6, 7\}$;\\
    sample $z_j \sim \mathrm{Unif}\{Z_1, Z_2, Z_3\}$, $\beta_0 \sim \mathrm{Unif}(0, 2)$, $\beta_1 \sim \mathrm{Unif}(-1, 1)$;\\
    $X_j \sim Binomial(n=C_j-1, p=\sigma(\beta_0 +\beta_1z_j))$.
    
    \item Ordinal Variables:\\
    For $j=1, \dots, 10$,\\
    set the ordinal level to be $K = \{3, 3, 3, 3, 4, 4, 4, 5, 5, 5\}$;\\
    sample $q_j \sim \mathrm{Unif}(n=K_j-1, 0, 1)$;\\
    sample $\omega_1 \sim \mathrm{Unif}(0.3, 0.9)$, $\omega_2 \sim \mathrm{Unif}(0.3, 0.9)$;\\
    $X_j = \bm{y}(Z_{1:2}, \bm{\omega}=(\omega_1, \omega_2)^T, K=K_j, \bm{q}=(q^{(1)}_j, \dots, q^{(K_j-1)}_j)^T)$, where $q^{(i)}_j$ is the sorted value indicated by order statistics.
    
    \item Continuous Variables:\\
    For $j=1, \dots, 10$,\\
    sample $\beta_0 \sim \mathrm{Unif}(1, 3)$, $\beta_1 \sim \mathrm{Unif}(1, 3)$;\\
    $$
    X_j =
    \begin{cases}
    \beta_0(Z_1 + Z_2) + \epsilon, & j \bmod 2 = 0,\\
    \beta_1(Z_1 Z_3) + \epsilon, & j \bmod 2 = 1.
    \end{cases}
    $$
\end{itemize}

\textbf{The distribution setting for $f_2$}
\begin{itemize}
   \item $Z \sim N(\bm{\mu} = (2, 3, 4)^T, \bm{\Sigma} = (d=1, \rho=0.8))$.
    \item Binary Variables:\\
    For $j = 1, \dots,50$,\\
    sample $z_j \sim \mathrm{Unif}\{Z_1, Z_2, Z_3\}$, $\beta_0 \sim \mathrm{Unif}(-1.5, 1.5)$, $\beta_1 \sim \mathrm{Unif}(-0.2, 0.8)$;\\
    $X_j \sim Binomial(n=1, p=\sigma(\beta_0 +\beta_1z_j))$.

    \item Categorical Variables:\\
    For $j=1, \dots, 5$,\\
    set the categorical level to be $C = \{3, 4, 5, 6, 7\}$;\\
    sample $z_j \sim \mathrm{Unif}\{Z_1, Z_2, Z_3\}$, $\beta_0 \sim \mathrm{Unif}(0, 4)$, $\beta_1=-1.5$;\\
    $X_j \sim Binomial(n=C_j-1, p=\sigma(\beta_0 +\beta_1z_j))$.
    
    \item Ordinal Variables:\\
    For $j=1, \dots, 10$,\\
    set the ordinal level to be $K = \{3, 3, 3, 3, 4, 4, 4, 5, 5, 5\}$;\\
    sample $q_j \sim \mathrm{Unif}(n=K_j-1, 0, 1)$;\\
    sample $\omega_1 \sim \mathrm{Unif}(0.3, 0.9)$, $\omega_2 \sim \mathrm{Unif}(0.3, 0.9)$;\\
    $X_j = \bm{y}(Z_{1:2}, \bm{\omega}=(\omega_1, \omega_2)^T, K=K_j, \bm{q}=(q^{(1)}_j, \dots, q^{(K_j-1)}_j)^T)$, where $q^{(i)}_j$ is the sorted value indicated by order statistics.
    
    \item Continuous Variables:\\
    For $j=1, \dots, 10$,\\
    sample $\beta_0 \sim \mathrm{Unif}(2, 4)$, $\beta_1 \sim \mathrm{Unif}(2, 4)$;\\
    $$
    X_j =
    \begin{cases}
    \beta_0(Z_1 + Z_2) + \epsilon, & j \bmod 2 = 0,\\
    \beta_1(Z_1 Z_3) + \epsilon, & j \bmod 2 = 1.
    \end{cases}
    $$
\end{itemize}

\textbf{The distribution setting for $f_3$}

\begin{itemize}
   \item $Z \sim N(\bm{\mu} = (3, 5, 7)^T, \bm{\Sigma} = (d=1, \rho=0.8))$.
    \item Binary Variables:\\
    For $j = 1, \dots,50$,\\
    sample $z_j \sim \mathrm{Unif}\{Z_1, Z_2, Z_3\}$, $\beta_0 \sim \mathrm{Unif}(-1.5, 1.5)$, $\beta_1 \sim \mathrm{Unif}(-1, 0.2)$;\\
    $X_j \sim Binomial(n=1, p=\sigma(\beta_0 +\beta_1z_j))$.

    \item Categorical Variables:\\
    For $j=1, \dots, 5$,\\
    set the categorical level to be $C = \{3, 4, 5, 6, 7\}$;\\
    sample $z_j \sim \mathrm{Unif}\{Z_1, Z_2, Z_3\}$, $\beta_0 \sim \mathrm{Unif}(2, 6)$, $\beta_1=-2$;\\
    $X_j \sim Binomial(n=C_j-1, p=\sigma(\beta_0 +\beta_1z_j))$.
    
    \item Ordinal Variables:\\
    For $j=1, \dots, 10$,\\
    set the ordinal level to be $K = \{3, 3, 3, 3, 4, 4, 4, 5, 5, 5\}$;\\
    sample $q_j \sim \mathrm{Unif}(n=K_j-1, 0, 1)$;\\
    sample $\omega_1 \sim \mathrm{Unif}(0.7, 0.9)$, $\omega_2 \sim \mathrm{Unif}(0.3, 0.5)$;\\
    $X_j = \bm{y}(Z_{1:2}, \bm{\omega}=(\omega_1, \omega_2)^T, K=K_j, \bm{q}=(q^{(1)}_j, \dots, q^{(K_j-1)}_j)^T)$, where $q^{(i)}_j$ is the sorted value indicated by order statistics.
    
    \item Continuous Variables:\\
    For $j=1, \dots, 10$,\\
    sample $\beta_0 \sim \mathrm{Unif}(-3, 1)$, $\beta_1 \sim \mathrm{Unif}(-3, 1)$;\\
    $$
    X_j =
    \begin{cases}
    \beta_0(Z_1 + Z_2) + \epsilon, & j \bmod 2 = 0,\\
    \beta_1(Z_1 Z_3) + \epsilon, & j \bmod 2 = 1.
    \end{cases}
    $$
\end{itemize}

We do experiment on treatment group with $n_{f_1}=2000$, big database with $n_{f_1}=20,000$, $n_{f_2}=80,000$, and $n_{f_3}=50,000$.

\begin{figure}[htbp]
    \centering
    \includegraphics[width=1.0\linewidth]{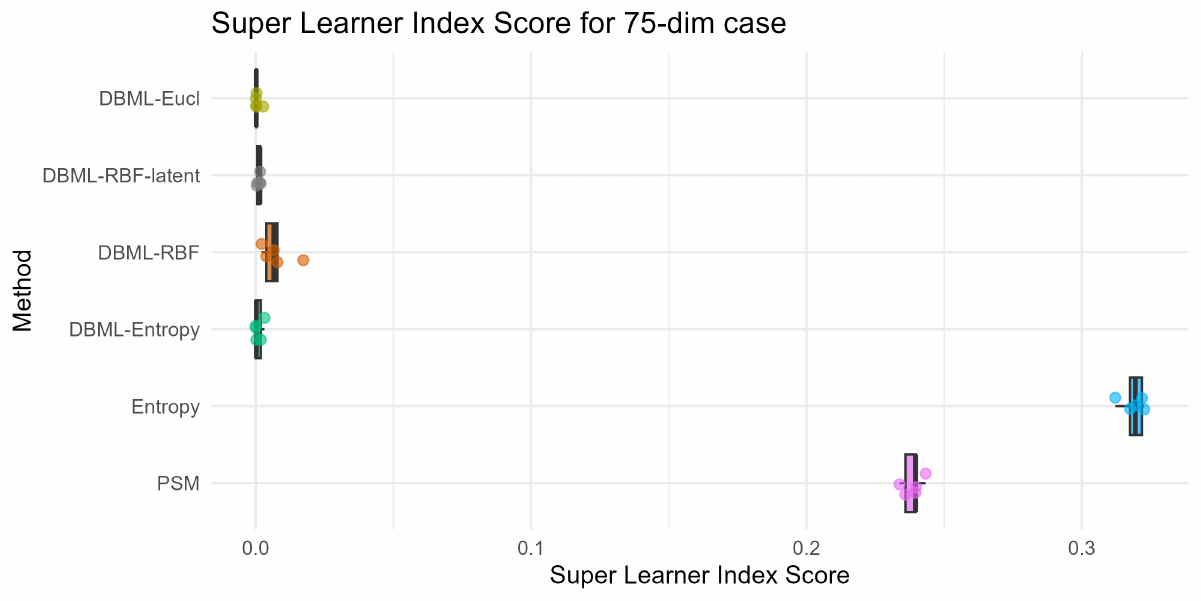}
    \caption{Boxplot for super learner score camparing different metrics: The scatter represents the actual value.}
    \label{fig:ex_sl}
\end{figure}

\begin{figure}[H]
    \centering
    \includegraphics[width=\linewidth]{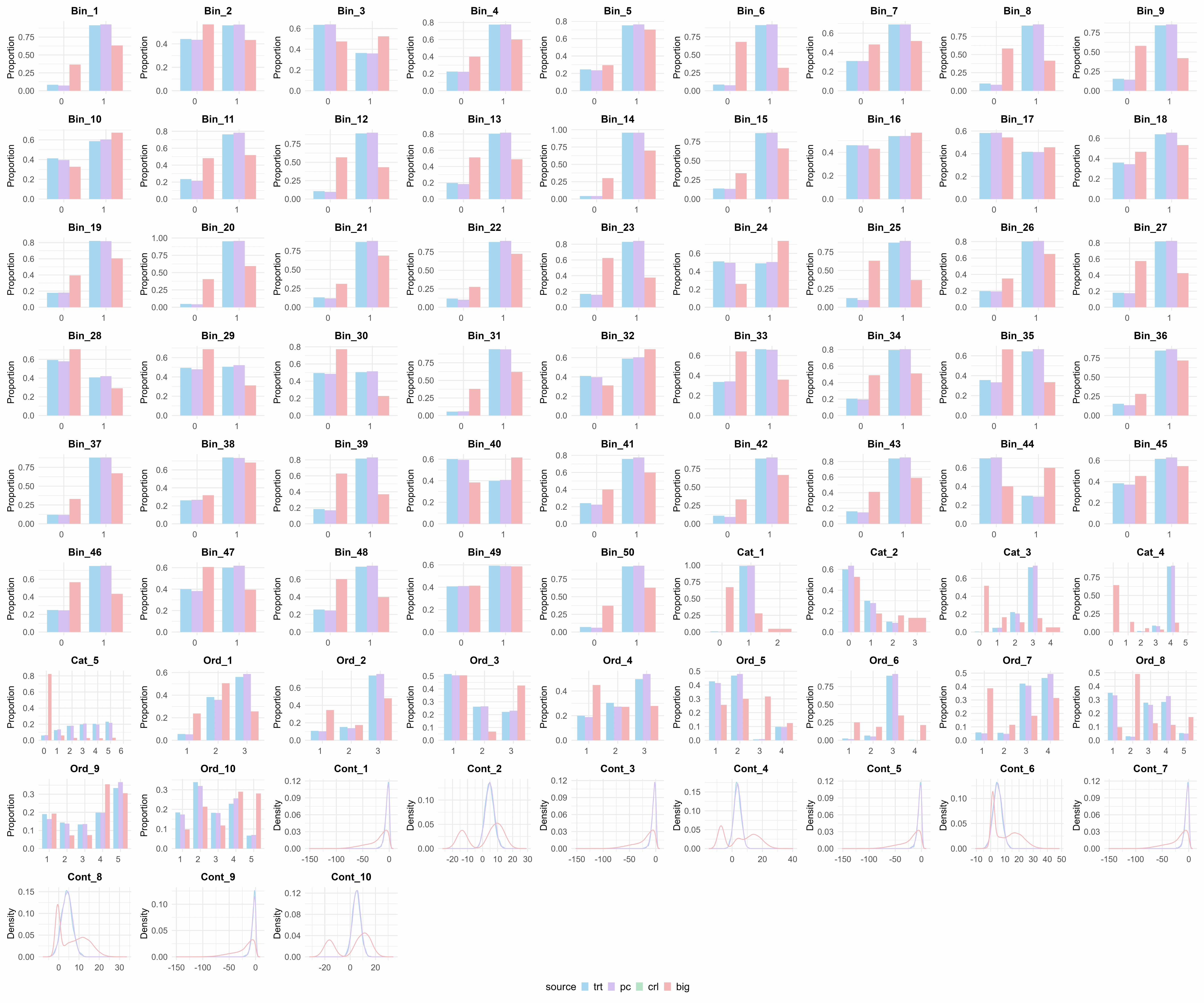}
    \caption{Distribution plot for each variables for 75-dim case: Comparing big database, possible controls after anomaly detection and treatment group.}
    \label{fig:dist75-1}
\end{figure}

\begin{figure}[H]
    \centering
    \includegraphics[width=\linewidth]{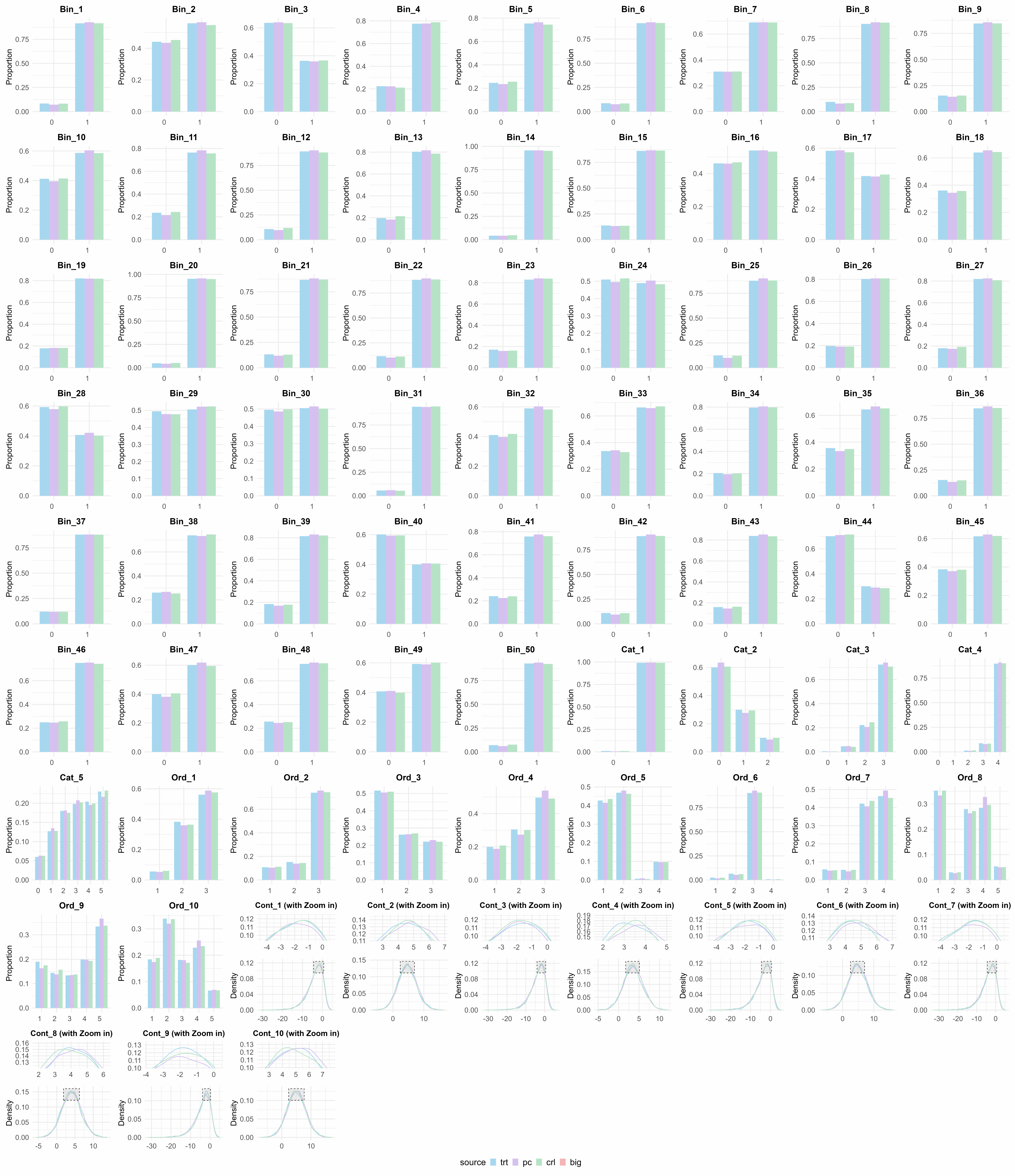}
    \caption{Distribution plot for each variables for 75-dim case: Comparing possible controls after anomaly detection, the final selected control group treatment group.}
    \label{fig:dist75-2}
\end{figure}

\begin{figure}[htbp]
    \centering
    \includegraphics[width=\linewidth]{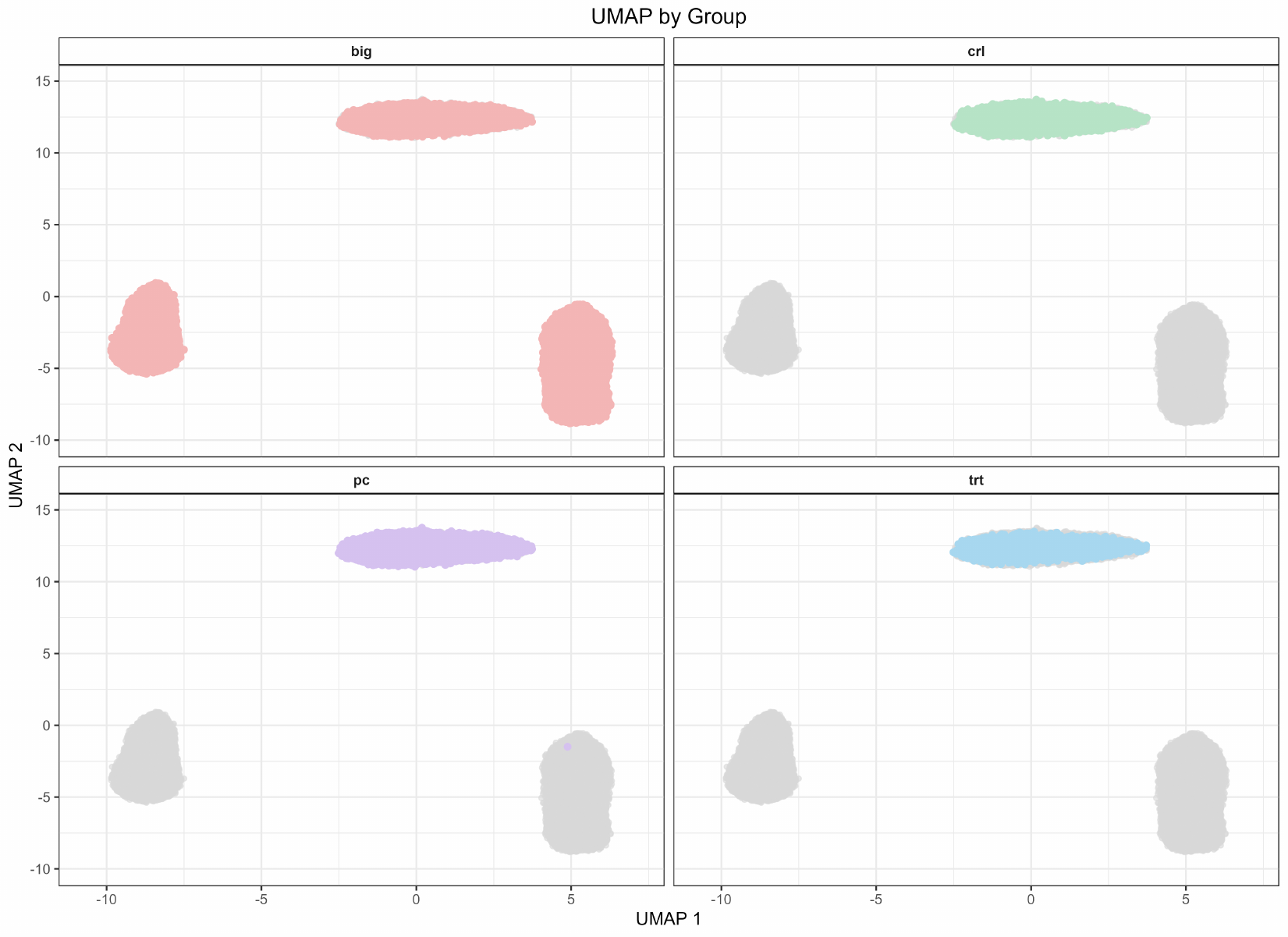}
    \caption{UMAP for 75-dim case: Comparing big database, possible controls, final selected control and treatment group.}
    \label{fig:umap75}
\end{figure}
The local neighborhood size was set to 100, proximities were calculated using the cosine distance metric, and the minimum distance was set to 0.25.

\section{Simulation Data Setting}
\label{ap:simulation}
Define $\sigma(\cdot)$ as the Standard Logistic Distribution function:
$$\sigma(x) = \frac{1}{1+e^{-x}}$$
\subsection{10-dimensional covariates}

\textbf{We first describe the distribution setting for $f_1$:}

\begin{itemize}
    \item $Z \sim N(\bm{\mu} = \mu_1, \bm{\Sigma} = (d=1, \rho))$, where $d$ indicates diagonal elements and $\rho$ indicates the off-diagonal elements.
    
    \item Binary/Categorical Variables:\\
    $X_1 \sim Binomial(n=1, p=\sigma(-1+0.1Z_1))$;\\
    $X_2 \sim Binomial(n=1, p=\sigma(0.3Z_1))$;\\
    $X_3 \sim Binomial(n=1, p=\sigma(-1+Z_1))$;\\
    $X_4 \sim Binomial(n=1, p=\sigma(-1+0.1Z_1))$;\\
    $X_5 \sim Binomial(n=1, p=\sigma(2+0.5Z_1))$;\\
    $X_6 \sim Binomial(n=3, p=\sigma(-0.1 +3Z_1))$; if $X_6=3$, set $X_6=1$.\\
    $X_7 \sim Binomial((n=5, p=\sigma(-0.8 + 0.2\log(5+Z_1)))$;  if $X_3=4$, set $X_3=2$;  if $X_3=5$, set $X_3=3$.\\
    Categorical Variables will be pre-processed as dummy variables.
    
    \item Ordinal Variables:\\
    Define function $\bm{y}(Z, \bm{\omega}, K, \bm{q})$ as following:\\
    Define $\mathbf{Y}^* = Z \bm{\omega} + \bm{\epsilon}$, where $\bm{\epsilon} \sim N(0,1)$.\\
    Let $\bm{\theta} = quantile(\mathbf{y}^*, \bm{q})$.\\
    Then define the ordinal output $y_i \in \{1, 2, \ldots, K\}$ for each sample $ i = 1, \ldots, n$ as:
    $$
    y_i = 1 + \sum_{j=1}^{K-1} \mathbb{I}(y_i^* > \theta_j)
    $$\\
    $X_8 = \bm{y}(Z_{1:2}, \bm{\omega}=(0.8,0.5)^T, K=4, \bm{q}=(0.25, 0.5, 0.75)^T)$;\\
    $X_9 = \bm{y}(Z_{1:2}, \bm{\omega}=(0.5,0.8)^T, K=4, \bm{q}=(0.1, 0.2, 0.3)^T)$;\\
   
    \item Continuous Variables:\\
    $X_{10}=Z_1Z_3+\bm{\epsilon}$.
\end{itemize}

\textbf{The distribution setting for $f_2$:}

\begin{itemize}
    \item $Z \sim N(\bm{\mu} = \mu_2, \bm{\Sigma} = (d=1, \rho))$.
    \item Binary/Categorical Variables:\\
    $X_1 \sim Binomial(n=1, p=\sigma(-1+0.3Z_1))$;\\
    $X_2 \sim Binomial(n=1, p=\sigma(0.1Z_1))$;\\
    $X_3 \sim Binomial(n=1, p=\sigma(-0.3+6Z_1))$;\\
    $X_4 \sim Binomial(n=1, p=\sigma(-1+0.4Z_1))$;\\
    $X_5 \sim Binomial(n=1, p=\sigma(0.8+0.5Z_1))$;\\
    $X_6 \sim Binomial(n=3, p=\sigma(-0.5 +0.6Z))$;\\
    $X_7 \sim Binomial((n=5, p=\sigma(-0.4 + 0.2\exp(0.5Z)))$.

    \item Ordinal Variables:\\
    $X_8 = \bm{y}(Z_{1:2}, \bm{\omega}=(0.1,0.5)^T, K=4, \bm{q}=(0.2, 0.4, 0.6)^T)$;\\
    $X_9 = \bm{y}(Z_{1:2}, \bm{\omega}=(0.2,0.9)^T, K=4, \bm{q}=(0.2, 0.4, 0.6)^T)$.
    \item Continuous Variables:\\
Define function $r(X, a, b)$ as following:\\
Define $\tilde{x} = \{ x_i \in x | a \leq x_i \leq b\}$,\\
\begin{align*}
    r(X, a, b) = 
    \begin{cases}
        \text{Uniform sample of size } |x| 
        \text{ with replacement } &\text{if } |\tilde{x}| < |x| \\
        x &\text{if } |\tilde{x}| = |x|
    \end{cases}
\end{align*}

    $X_{10}=r(5Z_1Z_3+\bm{\epsilon}, -5, 33)$.
\end{itemize}

\textbf{The distribution setting for $f_3$:}

\begin{itemize}
    \item $Z \sim N(\bm{\mu} = \mu_3, \bm{\Sigma} = (d=1, \rho))$.
    \item Binary/Categorical Variables:\\
    $X_1 \sim Binomial(n=1, p=\sigma(-1+0.3Z_1))$;\\
    $X_2 \sim Binomial(n=1, p=\sigma(0.1Z_1))$;\\
    $X_3 \sim Binomial(n=1, p=\sigma(-0.3+6Z_1))$;\\
    $X_4 \sim Binomial(n=1, p=\sigma(-1+0.4Z_1))$;\\
    $X_5 \sim Binomial(n=1, p=\sigma(0.8+0.5Z_1))$;\\
    $X_6 \sim Binomial(n=3, p=\sigma(-0.5 +0.6Z))$;\\
    $X_7 \sim Binomial((n=5, p=\sigma(-0.4 + 0.2\exp(0.5Z)))$.
    
    \item Ordinal Variables:\\
    $X_8 = \bm{y}(Z_{1:2}, \bm{\omega}=(0.1,0.5)^T, K=4, \bm{q}=(0.5, 0.8, 0.9)^T)$;\\
    $X_9 = \bm{y}(Z_{1:2}, \bm{\omega}=(0.2,0.9)^T, K=4, \bm{q}=(0.8, 0.85, 0.9)^T)$.

    \item Continuous Variables:\\
     $X_{10}=5Z_1Z_3+\bm{\epsilon}$.
\end{itemize}

\subsection{20-dimensional covariates}

\textbf{The distribution setting for $f_1$}
\begin{itemize}
   \item $Z \sim N(\bm{\mu} = \mu_1, \bm{\Sigma} = (d=1, \rho))$
    
    \item Binary/Categorical Variables:\\
    $X_1 \sim Binomial(n=1, p=\sigma(-1+0.1Z_1))$;\\
    $X_2 \sim Binomial(n=1, p=\sigma(0.3Z_1))$;\\
    $X_3 \sim Binomial(n=1, p=\sigma(-3+Z_1))$;\\
    $X_4 \sim Binomial(n=1, p=\sigma(-1+0.1Z_2))$;\\
    $X_5 \sim Binomial(n=1, p=\sigma(2+0.5Z_2))$;\\
    $X_6 \sim Binomial(n=1, p=\sigma(-4-0.1Z_2))$;\\
    $X_7 \sim Binomial(n=1, p=\sigma(2-0.2Z_3))$;\\
    $X_8 \sim Binomial(n=1, p=\sigma(-3-0.2Z_3))$;\\
    $X_9 \sim Binomial(n=2, p=\sigma(-0.1 +3Z_1))$; if $X_9=2$, set $X_9=1$.\\
    $X_{10} \sim Binomial((n=4, p=\sigma(-0.8 + 0.2\log(5+Z_2)))$;  if $X_{10}=3$, set $X_{10}=0$;  if $X_{10}=4$, set $X_{10}=2$.\\
    $X_{11} \sim Binomial(n=3, p=\sigma(-2+0.6Z_2))$;\\
    $X_{12} \sim Binomial(n=3, p=\sigma(0.8+0.1\exp(Z_1)))$.

    \item Ordinal Variables:\\
    $X_{13} = \bm{y}(Z_{1:2}, \bm{\omega}=(1, 0)^T, K=4, \bm{q}=(0.25, 0.5, 0.75)^T)$;\\
    $X_{14} = \bm{y}(Z_{1:2}, \bm{\omega}=(0, 1)^T, K=4, \bm{q}=(0.25, 0.5, 0.75)^T)$;\\
    $X_{15} = \bm{y}(Z_{1:2}, \bm{\omega}=(0.5, 0.8)^T, K=4, \bm{q}=(0.1, 0.2, 0.3)^T)$;\\
    $X_{16} = \bm{y}(Z_{1:2}, \bm{\omega}=(0.8, 0.5)^T, K=4, \bm{q}=(0.3, 0.4, 0.5)^T)$;\\
    $X_{17} = \bm{y}(Z_{1:2}, \bm{\omega}=(1, 1)^T, K=4, \bm{q}=(0.6, 0.7, 0.8)^T)$.

    \item Continuous Variables:\\
    $X_{18} = \cos(Z_3) + \epsilon$;\\
    $X_{19} = Z_1 + Z_3 +\epsilon$;\\
    $X_{20} = Z_1Z_3 + \epsilon$.
\end{itemize}

\textbf{The distribution setting for $f_2$}
\begin{itemize}
   \item $Z \sim N(\bm{\mu} = \mu_2, \bm{\Sigma} = (d=1, \rho))$
    
    \item Binary/Categorical Variables:\\
    $X_1 \sim Binomial(n=1, p=\sigma(-1+0.3Z_1))$;\\
    $X_2 \sim Binomial(n=1, p=\sigma(0.1Z_1))$;\\
    $X_3 \sim Binomial(n=1, p=\sigma(-0.3+6Z_1))$;\\
    $X_4 \sim Binomial(n=1, p=\sigma(-1+0.4Z_2))$;\\
    $X_5 \sim Binomial(n=1, p=\sigma(0.8+0.5Z_2))$;\\
    $X_6 \sim Binomial(n=1, p=\sigma(-4+0.1Z_2))$;\\
    $X_7 \sim Binomial(n=1, p=\sigma(0.2-0.6Z_3))$;\\
    $X_8 \sim Binomial(n=1, p=\sigma(-2+0.2Z_3))$;\\
    $X_9 \sim Binomial(n=2, p=\sigma(-0.5 +0.6Z_1))$; if $X_9=2$, set $X_9=1$.\\
    $X_{10} \sim Binomial((n=4, p=\sigma(-0.4 + 0.2\exp(0.5Z_2)))$;  if $X_{10}=3$, set $X_{10}=0$;  if $X_{10}=4$, set $X_{10}=2$.\\
    $X_{11} \sim Binomial(n=3, p=\sigma(-2+0.1Z_2))$;\\
    $X_{12} \sim Binomial(n=3, p=\sigma(0.5+0.8\exp(Z_1)))$.

    \item Ordinal Variables:\\
    $X_{13} = \bm{y}(Z_{1:2}, \bm{\omega}=(0.9, 0)^T, K=4, \bm{q}=(0.1, 0.3, 0.5)^T)$;\\
    $X_{14} = \bm{y}(Z_{1:2}, \bm{\omega}=(0, 0.8)^T, K=4, \bm{q}=(0.1, 0.3, 0.5)^T)$;\\
    $X_{15} = \bm{y}(Z_{1:2}, \bm{\omega}=(0.1, 0.5)^T, K=4, \bm{q}=(0.2, 0.4, 0.6)^T)$;\\
    $X_{16} = \bm{y}(Z_{1:2}, \bm{\omega}=(0.2, 0.9)^T, K=4, \bm{q}=(0.2, 0.4, 0.6)^T)$;\\
    $X_{17} = \bm{y}(Z_{1:2}, \bm{\omega}=(1, -1)^T, K=4, \bm{q}=(0.2, 0.4, 0.6)^T)$.

    \item Continuous Variables:\\
    $X_{18} = r(\sin(Z_3) + \epsilon, -2, 4)$;\\
    $X_{19} = r(Z_1 - Z_3 +\epsilon, -4, 13)$;\\
    $X_{20} = r(5Z_1Z_3 + \epsilon, -5, 33)$.
\end{itemize}

\textbf{The distribution setting for $f_3$}

\begin{itemize}
   \item $Z \sim N(\bm{\mu} = \mu_3, \bm{\Sigma} = (d=1, \rho))$
    
    \item Binary/Categorical Variables:\\
    $X_1 \sim Binomial(n=1, p=\sigma(-1+0.3Z_1))$;\\
    $X_2 \sim Binomial(n=1, p=\sigma(0.1Z_1))$;\\
    $X_3 \sim Binomial(n=1, p=\sigma(-0.3+6Z_1))$;\\
    $X_4 \sim Binomial(n=1, p=\sigma(-1+0.4Z_2))$;\\
    $X_5 \sim Binomial(n=1, p=\sigma(0.8+0.5Z_2))$;\\
    $X_6 \sim Binomial(n=1, p=\sigma(-4+0.1Z_2))$;\\
    $X_7 \sim Binomial(n=1, p=\sigma(0.2-0.6Z_3))$;\\
    $X_8 \sim Binomial(n=1, p=\sigma(-2+0.2Z_3))$;\\
    $X_9 \sim Binomial(n=2, p=\sigma(-0.5 +0.6Z_1))$;\\
    $X_{10} \sim Binomial((n=4, p=\sigma(-0.4 + 0.2\exp(0.5Z_2)))$;\\
    $X_{11} \sim Binomial(n=3, p=\sigma(-4+Z_2^2))$;\\
    $X_{12} \sim Binomial(n=3, p=\sigma(-1.5+0.8\exp(2Z_1)))$.

    \item Ordinal Variables:\\
    $X_{13} = \bm{y}(Z_{1:2}, \bm{\omega}=(0.9, 0)^T, K=4, \bm{q}=(0.1, 0.3, 0.5)^T)$;\\
    $X_{14} = \bm{y}(Z_{1:2}, \bm{\omega}=(0, 0.8)^T, K=4, \bm{q}=(0.1, 0.3, 0.5)^T)$;\\
    $X_{15} = \bm{y}(Z_{1:2}, \bm{\omega}=(0.1, 0.5)^T, K=4, \bm{q}=(0.2, 0.4, 0.6)^T)$;\\
    $X_{16} = \bm{y}(Z_{1:2}, \bm{\omega}=(0.2, 0.9)^T, K=4, \bm{q}=(0.8, 0.85, 0.9)^T)$;\\
    $X_{17} = \bm{y}(Z_{1:2}, \bm{\omega}=(1, -1)^T, K=4, \bm{q}=(0.8, 0.85, 0.9)^T)$.

    \item Continuous Variables:\\
    $X_{18} = \sin(Z_2)+\epsilon$;\\
    $X_{19} = Z_1 - Z_3 + \epsilon$;\\
    $X_{20} = 5Z_1Z_3 +\epsilon$.
\end{itemize}

\subsection{Real-world databases}
\label{ap:big}
The data generation process for treatment groups and large real-world databases based on $f_1, f_2$ and $f_3$ is as follows:
\begin{itemize}
    \item s: Treatment group contains 500 units with $X \sim f_1$.
    \item S: Treatment group contains 2000 units with $X \sim f_1$.

    \item Big real-world database D contains $150,000$ units with:
    \begin{equation*}
        \begin{cases}
             n_{f_1} &=20,000\\
             n_{f_2} &=80,000\\
             n_{f_3} &=50,000
        \end{cases}
    \end{equation*}
    \item Big real-world database d contains $150,000$ units with:
    \begin{equation*}
        \begin{cases}
             n_{f_1} &=40,000\\
             n_{f_2} &=60,000\\
             n_{f_3} &=50,000
        \end{cases}
    \end{equation*}
\end{itemize}

\subsection{Hierarchical Repeating Experiment Simulation}
We set the simulation study experiment via a hierarchical structure that allows varying parameters in different ranges. We consider (i) the number of units in treatment group, (ii) Correlation of covariates $\rho$, (iii) Proportion of different distributions in the big real-world dataset, (iv) Dimension of covariates, and with these for aspects we can form a 2 factorial design like experiment setting. Also, to take randomness into account, we simulate 20 replication datasets for each of the $2^4=16$ design, and do both the local pivotal method selection and propensity score matching 5 times for each replication:

\begin{itemize}
    \item Number of units in treatment group: 500 or 2000 units.
    \item Correlation of covariates: high or low. For each replication, simulate $\mu_1 \sim N((1, 2, 3)^T, 0.1)$, $\mu_2 \sim N((2, 3, 4)^T, 0.1)$, $\mu_3 \sim N((3, 5, 7)^T, 0.1)$, $\rho \sim \text{Uniform}(0, 0.1)$ for low correlation design and $\rho \sim \text{Uniform}(0.8, 0.85)$ for high correlation design.
    \item Proportion of different distributions in the big real-world dataset: either the big real-world database 1 or 2 described in section \ref{ap:big}.
    \item Dimension of covariates: $p=10$ or 20.
\end{itemize}

\begin{table}[H]
\centering
\caption{Simulation factorial Design Experiment}
\label{tab:sim_results}
\begin{tabular}{ccccccc}
\toprule
\shortstack{Size of \\ Treatments} & \shortstack{Correlation of \\ latent variables} & \shortstack{Proportion of distinct \\ groups in big database} & \shortstack{Dimension of \\ Covariates} \\
\midrule
500 & $\rho \sim \text{Uniform}(0, 0.1)$ & 4:6:5 & 10  \\
\hline
500 & $\rho \sim \text{Uniform}(0, 0.1)$  & 4:6:5 & 20 \\
\hline
500 & $\rho \sim \text{Uniform}(0, 0.1)$  & 2:8:5 & 10 \\
\hline
500 & $\rho \sim \text{Uniform}(0, 0.1)$  & 2:8:5 & 20 \\
\hline
500 & $\rho \sim \text{Uniform}(0.8, 0.85)$  & 4:6:5 & 10 \\
\hline
500 & $\rho \sim \text{Uniform}(0.8, 0.85)$ & 4:6:5 & 20  \\
\hline
500 & $\rho \sim \text{Uniform}(0.8, 0.85)$ & 2:8:5 & 10  \\
\hline
500 & $\rho \sim \text{Uniform}(0.8, 0.85)$ & 2:8:5 & 20 \\
\hline
2000 & $\rho \sim \text{Uniform}(0, 0.1)$  & 4:6:5 & 10  \\
\hline
2000 & $\rho \sim \text{Uniform}(0, 0.1)$  & 4:6:5 & 20  \\
\hline
2000 & $\rho \sim \text{Uniform}(0, 0.1)$  & 2:8:5 & 10 \\
\hline
2000 & $\rho \sim \text{Uniform}(0, 0.1)$  & 2:8:5 & 20  \\
\hline
2000 & $\rho \sim \text{Uniform}(0.8, 0.85)$ & 4:6:5 & 10  \\
\hline
2000 & $\rho \sim \text{Uniform}(0.8, 0.85)$ & 4:6:5 & 20 \\
\hline
2000 & $\rho \sim \text{Uniform}(0.8, 0.85)$ & 2:8:5 & 10 \\
\hline
2000 & $\rho \sim \text{Uniform}(0.8, 0.85)$ & 2:8:5 & 20  \\
\bottomrule
\end{tabular}
\end{table}

\subsection{Correlation among variables}
In Table~\ref{tab:simulation-factors} and Table~\ref{tab:sim_results}, we mentioned that we are comparing two levels of the correlation among variables in our simulation design.  The correlation level is defined using the correlation parameter $\rho$ in the covariance matrix of the latent variable $Z$, with $Z$ following a multivariate Gaussian distribution.  In this appendix section we show the actual correlation among covariates $X$ for different settings.

To evaluate the pairwise associations and assess potential multicollinearity among our multi-type covariates,  we utilized the automated framework provided by the \texttt{correlation} package within the \texttt{easystats} ecosystem in R \cite{makowski2020methods, makowski2022correlation}. This algorithm dynamically selected the optimal correlation metric for each specific pair of data types: standard Pearson correlation for continuous-continuous pairs, point-biserial correlation for continuous-binary pairs, and tetrachoric or polychoric coefficients for categorical combinations. This holistic approach yields a mathematically rigorous, unified correlation matrix mapping all parameters onto a standardized scale ($r \in [-1, 1]$), ensuring a reliable and methodologically sound assessment of the covariate dependencies.

\begin{table}[H]
\centering
\caption{Summary of Absolute Correlation $|r|$ on the treatment group covariates}
\label{tab:correlation_metrics}
\begin{tabular}{lcccccc}
\toprule
\textbf{Design} & \textbf{Q1} & \textbf{median} & \textbf{mean} & \textbf{Q3} & \textbf{max} & \textbf{$\% |r| \ge 0.50$} \\
\midrule
      \texttt{SCDP} & 0.0115 & 0.0260 & 0.0633 & 0.0570 & 0.8194 & 1.83 \\
      \texttt{SCdP} & 0.0115 & 0.0260 & 0.0633 & 0.0570 & 0.8194 & 1.83 \\
      \texttt{sCDP} & 0.0205 & 0.0436 & 0.0768 & 0.0813 & 0.8340 & 1.86 \\
      \texttt{sCdP} & 0.0205 & 0.0436 & 0.0768 & 0.0813 & 0.8340 & 1.86 \\
      \texttt{ScDP} & 0.0092 & 0.0205 & 0.0464 & 0.0417 & 0.7688 & 0.77 \\
      \texttt{ScdP} & 0.0092 & 0.0205 & 0.0464 & 0.0417 & 0.7688 & 0.77 \\
      \texttt{scDP} & 0.0181 & 0.0377 & 0.0610 & 0.0678 & 0.7822 & 0.81 \\
      \texttt{scdP} & 0.0181 & 0.0377 & 0.0610 & 0.0678 & 0.7822 & 0.81 \\
      \texttt{SCDp} & 0.0161 & 0.0384 & 0.0989 & 0.1000 & 0.8820 & 4.00 \\
      \texttt{SCdp} & 0.0161 & 0.0384 & 0.0989 & 0.1000 & 0.8820 & 4.00 \\
      \texttt{sCDp} & 0.0232 & 0.0507 & 0.1071 & 0.1100 & 0.8918 & 4.00 \\
      \texttt{sCdp} & 0.0232 & 0.0507 & 0.1071 & 0.1100 & 0.8918 & 4.00 \\
      \texttt{ScDp} & 0.0171 & 0.0378 & 0.0870 & 0.0759 & 0.8816 & 3.36 \\
      \texttt{Scdp} & 0.0171 & 0.0378 & 0.0870 & 0.0759 & 0.8816 & 3.36 \\
      \texttt{scDp} & 0.0223 & 0.0476 & 0.0961 & 0.0918 & 0.9027 & 3.36 \\
      \texttt{scdp} & 0.0223 & 0.0476 & 0.0961 & 0.0918 & 0.9027 & 3.36 \\
      \bottomrule
\end{tabular}
\end{table}
According to Table~\ref{tab:correlation_metrics}, the correlation among covariates $X$ is not only determined by the correlation parameter $\rho$ of the latent variable $Z$, but also affected by the dimension and the sample size of the dataset. In our simulation study design, the data with lower dimension $p$ and smaller sample size $s$ tend to have higher absolute correlation $|r|$ among the covariates. Specifically, the correlation and the dimension shows strong correlated pattern.

\section{Algorithm Implementation Trick}
\label{sec:trick}
To implement the algorithm well, we need to focus on: 
\begin{itemize}
    \item Number of latent dimension selection: \\
    For a modality with $p>10$, choose the latent dimension starting from 3 and increase it if necessary. For a modality with $p<=10$, set the latent dimension close to $p/2$. For latent VAE, start from 3.
    
    \item Number of hidden dimensions in VAE: For first stage, use 50 as default, for the second stage use 32 as default. 
    
    \item Ordinal modality VAE embedding dimension: choose the highest value of $K$.
    
    \item Learning rate and schedule: Set the learning rate = 0.005 and use the cosine scheduler. For Kernel MMD running, set the learning rate=0.05 and decrease if necessary.
    
    \item Epoch: by default, 1000 for each VAE, and 50000 for KMMD. Check the VAE reconstruction accuracy of the treatment group to decide tuning hyper parameters.

    \item Anomaly Detection quantile: choose $q=1$ as default or higher to avoid the over detection.
\end{itemize}

\end{document}